\documentclass[]{aastex7}

\newcommand\lya{Ly$\alpha$}

\usepackage{rotating}
\usepackage{longtable}
\usepackage{xcolor}
\usepackage{graphicx}      
\usepackage{adjustbox}     
\usepackage{array} 
\usepackage{array} 
\newcommand{\rot}[1]{\rotatebox[origin=c]{45}{#1}} 
\usepackage{threeparttablex}
\usepackage{makecell}
\usepackage{hyperref}

\begin{document}


\title{HWO Target Stars and Systems: A Survey of Archival UV and X-ray Data}

\author[0000-0002-1046-025X]{Sarah Peacock} 
\affiliation{University of Maryland Baltimore County, Baltimore, MD, USA}
\affiliation{NASA Goddard Space Flight Center, Greenbelt, MD, USA}
\email{sarah.r.peacock@nasa.gov}

\author[0000-0001-9667-9449]{David J. Wilson}
\affil{Laboratory for Atmospheric and Space Physics, University of Colorado Boulder, Boulder, CO 80309}
\email{david.wilson@lasp.colorado.edu}

\author[0000-0003-1290-3621]{Tyler Richey-Yowell}
\affil{Lowell Observatory, 1400 W. Mars Hill Road, Flagstaff, AZ 86001, USA}
\affil{Percival Lowell Postdoctoral Fellow}
\email{try@lowell.edu}

\author[0000-0003-3989-5545]{Noah W. Tuchow}
\affiliation{NASA Goddard Space Flight Center, Greenbelt, MD, USA}
\email{noah.w.tuchow@nasa.gov}

\author[0000-0002-1002-3674]{Kevin France}
\affiliation{Laboratory for Atmospheric and Space Physics, University of Colorado Boulder, Boulder, CO 80309}
\email{kevin.france@colorado.edu}

\author[0000-0002-7349-1387]{Jos\'e A. Caballero} 
\affiliation{Centro de Astrobiolog\'ia, CSIC-INTA, 
	Camino Bajo del Castillo s/n, Campus ESAC, 
	28692 Villanueva de la Ca\~nada, Madrid, Spain}
\email{caballero@cab.inta-csic.es}

\author[0000-0001-5252-5042]{Riccardo Spinelli}
\affiliation{INAF – Osservatorio Astronomico di Palermo, Piazza del Parlamento 1, I-90134, Palermo, Italy}
\affiliation{ Como Lake Center for Astrophysics (CLAP), DiSAT, Univerist\`a degli Studi dell'Insubria, via Valleggio 11, I-22100 Como, Italy}
\email{riccardo.spinelli@inaf.it}

\author[0000-0002-5466-3817]{L\'ia Corrales} 
\affiliation{University of Michigan, Ann Arbor, MI 48109, USA} 
\email{liac@umich.edu}

\author[0000-0002-6387-7729]{Aiden S. Zelakiewicz}
\affiliation{Department of Astronomy and Carl Sagan Institute, Cornell University, 122 Sciences Drive, Ithaca, NY 14853, USA}
\email{asz39@cornell.edu}

\author[0000-0003-3786-3486]{Seth Redfield}
\affiliation{Astronomy Department and Van Vleck Observatory, Wesleyan University, Middletown, CT 06459, USA}
\email{sredfield@wesleyan.edu}

\author[0000-0003-1337-723X]{Keighley Rockcliffe} 
\affiliation{University of Maryland Baltimore County, Baltimore, MD, USA}
\affiliation{NASA Goddard Space Flight Center, Greenbelt, MD, USA}
\email{keighley.e.rockcliffe@nasa.gov}

\author[0000-0002-1176-3391]{Allison Youngblood}
\affiliation{NASA Goddard Space Flight Center, Greenbelt, MD, USA}
\email{allison.a.youngblood@nasa.gov}

\author[0000-0001-8499-2892]{Cynthia S. Froning}
\affiliation{Southwest Research Institute, San Antonio, TX 78238}
\email{cynthia.froning@swri.org}

\author[0000-0002-7119-2543]{Girish M. Duvvuri}
\affiliation{Department of Physics and Astronomy, Vanderbilt University, Nashville, TN 37235, USA}
\email{girish.duvvuri@vanderbilt.edu}

\author[0000-0002-4955-0471]{Breanna A. Binder}
\affiliation{Department of Physics and Astronomy, California State Polytechnic University, Pomona, CA, USA}
\email{babinder@cpp.edu}

\author[0000-0003-0595-5132]{Natalie R. Hinkel}
\affiliation{Department of Physics \& Astronomy, Louisiana State University, 202 Nicholson Hall Baton Rouge, LA 70803}
\email{natalie.hinkel@gmail.com}

\author[0000-0003-2008-1488]{Eric E.\ Mamajek}
\affiliation{Jet Propulsion Laboratory, California Institute of Technology, 4800 Oak Grove Dr., Pasadena, CA 91109, USA}
\email{mamajek@jpl.nasa.gov}

\begin{abstract}
We assess archival high-energy data for key stars on the Habitable Worlds Observatory (HWO) Target Stars and Systems 2025 list, as stellar radiation is critical to shaping and interpreting planetary atmospheres. Using a sample of 98 nearby stars (HWO Tier 1 targets), we compile and evaluate X-ray and ultraviolet (UV) data from archival eROSITA, Chandra, XMM-Newton, ROSAT, EUVE, Swift, FUSE, IUE, GALEX, and HST. We examine spectral and temporal coverage, assess data quality, and identify major gaps. UV data are moderately available, with most coverage coming from near-UV spectra from IUE. Far fewer stars have far-UV spectra, especially from HST. In the X-ray regime, some stars have high-quality spectra, while others are limited to shallow detections or broad-band photometry. A small fraction of the sample has both X-ray and UV spectra of sufficient quality to support full spectral energy distribution modeling. Truly comprehensive coverage across X-ray, extreme-UV, and both UV bands remains extremely rare. Most datasets are single-epoch, limiting assessments of variability and flares - key factors in atmospheric photochemistry and escape. Moreover, the lack of simultaneous or contemporaneous observations across bands adds further uncertainty. Our findings underscore the need for new space-based missions and coordinated multiwavelength campaigns, ideally with overlapping coverage, to improve stellar characterization for HWO. As several key observatories age and face potential decommissioning, there is a narrow window of opportunity to secure these critical data. Investing in this effort now will directly support the science goals of HWO and enhance future studies of planetary habitability.

\end{abstract}


\section{Introduction} 

NASA's Habitable Worlds Observatory (HWO) is being formulated to achieve the goal of direct imaging and spectral characterization of potentially habitable worlds in the habitable zones of nearby Sun-like stars. Its primary goal, often used as a benchmark for mission success, is to detect and spectroscopically characterize at least 25 potentially habitable exoplanets \citep[e.g.][]{Feinberg2024}. In support of mission planning, NASA established dedicated science and technology working groups. The science working groups are charged with defining HWO's science objectives, translating 2020 Astrophysics Decadal Survey \citep{astro2020} recommendations into concrete mission requirements, and identifying supporting data needs. The technology working groups, while not the focus of this paper, assess mission architectures and the technical capabilities required to achieve these goals
\citep[e.g.][]{Feinberg2024,Stark2024,Mennesson2024}.

To maximize the scientific return of HWO, especially in its search for biosignatures, a robust understanding of the high-energy radiation environment of target stars is essential and inherently a multi-wavelength characterization, given the various manifestations of high-energy emission across spectral types. This includes ultraviolet (UV) and X-ray fluxes that regulate atmospheric chemistry, drive escape processes, and influence the detectability of biological markers. Stellar high-energy radiation (spanning approximately 1 to 3200\,\AA) plays a central role in shaping planetary atmospheres and interpreting their observable properties. This regime comprises the X-ray ($<$100\,\AA), extreme-UV (EUV; 100--912\,\AA), far-UV (FUV; 912--1800\,\AA), and near-UV (NUV; 1800--3200\,\AA) bands. Emitted from the chromosphere, transition region, and corona, these fluxes are manifestations of magnetic activity and exhibit variability as a result of flares, spots, and stellar cycles. Each spectral sub-region contributes distinctly to planetary atmospheric evolution and potential habitability:
\begin{itemize}
    \item \textbf{NUV radiation} (1800--3200\,\AA) drives key photochemical reactions, particularly affecting O$_2$ and O$_3$ stability \citep[e.g.,][]{Seager2000, segura2010, Hu2012, Harman2015, tilley2019}, while simultaneously posing a biological hazard by degrading nucleic acids and other biomolecules and depopulating metastable helium that is used as an observation indicator for atmospheric escape \citep{sagan1973,karentz1991, cockell1998, Oklopcic2019}.
    \item \textbf{FUV radiation} (912--1800\,\AA) photodissociates major molecules such as H$_2$O, CO$_2$, H$_2$, and CH$_4$, initiating haze formation in reducing atmospheres \citep[e.g.,][]{trainer2006, Zerkle2012,Arney2017}. Lyman-$\alpha$ (\lya) emission at 1216 \AA, which alone can dominate FUV radiation for cool stars, has also been implicated in the photoproduction of amino acids \citep{Bernstein2002,MunozCaro2002,MunozCaro2014}. Hazes also contribute to absorption and reflection of the atmosphere, altering climate, in addition to obscuring molecular features from detection \citep[e.g.,][]{Arney2017, Arney2018}.
    \item \textbf{EUV radiation} (100--912\,\AA) dominates upper atmospheric (ionosphere, exosphere, and thermosphere) heating and ionization, leading to atmospheric expansion and escape \citep[e.g.,][]{lammer2007,murrayclay2009, koskinen2010,Chadney2015,Johnstone2019}.
    \item \textbf{X-rays} ($<$100\,\AA) penetrate deeper atmospheric layers (thermosphere and top of mesosphere) and contribute significantly to thermal escape processes on giant planets \citep[e.g.,][]{lopez2012, owen2012}.
\end{itemize}

Because Earth's atmosphere is opaque to UV and X-rays, direct observations of high-energy stellar radiation rely on space-based telescopes. Some of these observations are further complicated by absorption from the local interstellar medium \citep[LISM, ][]{Frisch2011}, which can strongly attenuate EUV photons and specific UV lines, particularly \lya\ and the Mg~{\sc ii} h\&k doublet. Accurate reconstruction of intrinsic stellar emission thus requires knowledge of ISM column densities along each line of sight, especially for HWO's anticipated targets within $\sim$1--25\,pc \citep{Wood2005,Youngblood2025}.

Recent planning efforts by HWO science working groups, including the \textit{Exoplanet Science Yield} (linking science goals to mission designs for yield predictions) and \textit{Living Worlds} (investigating the observations required to detect life on other planets) groups, have highlighted several critical stellar high-energy data needs to ensure mission success. These include the supplying of narrowband NUV fluxes to refine coronagraph exposure-time calculations, the acquisition of representative UV spectra to use as input for photochemical models of exoplanet atmospheres, and the characterization of ISM absorption features to recover intrinsic stellar line strengths. In addition, improving reconstructions of EUV flux through direct observations and empirical correlations with FUV diagnostics and X-ray measurements is essential for estimating atmospheric escape and habitability. The panel report from NASA's 2025 Senior Review of Operating Missions also cited the need for UV and X-ray data to estimate stellar EUV flux as a motivation for the continued operation of the Hubble Space Telescope (HST), the Chandra X-ray Observatory, the Neil Gehrels Swift Observatory (Swift), and the X-ray Multi-Mirror mission (XMM-Newton). For late-type stars in particular, where empirical UV data are sparse and model uncertainties are high, expanding the combined X-ray and EUV (XUV; 1--912 \AA) dataset is vital to inform target selection, assess biosignature detectability, and evaluate planetary stability in the face of stellar activity \citep{Greene2023, Weiner2024}.

\begin{figure}[ht!]
    \centering
    \includegraphics[width=0.85\linewidth]{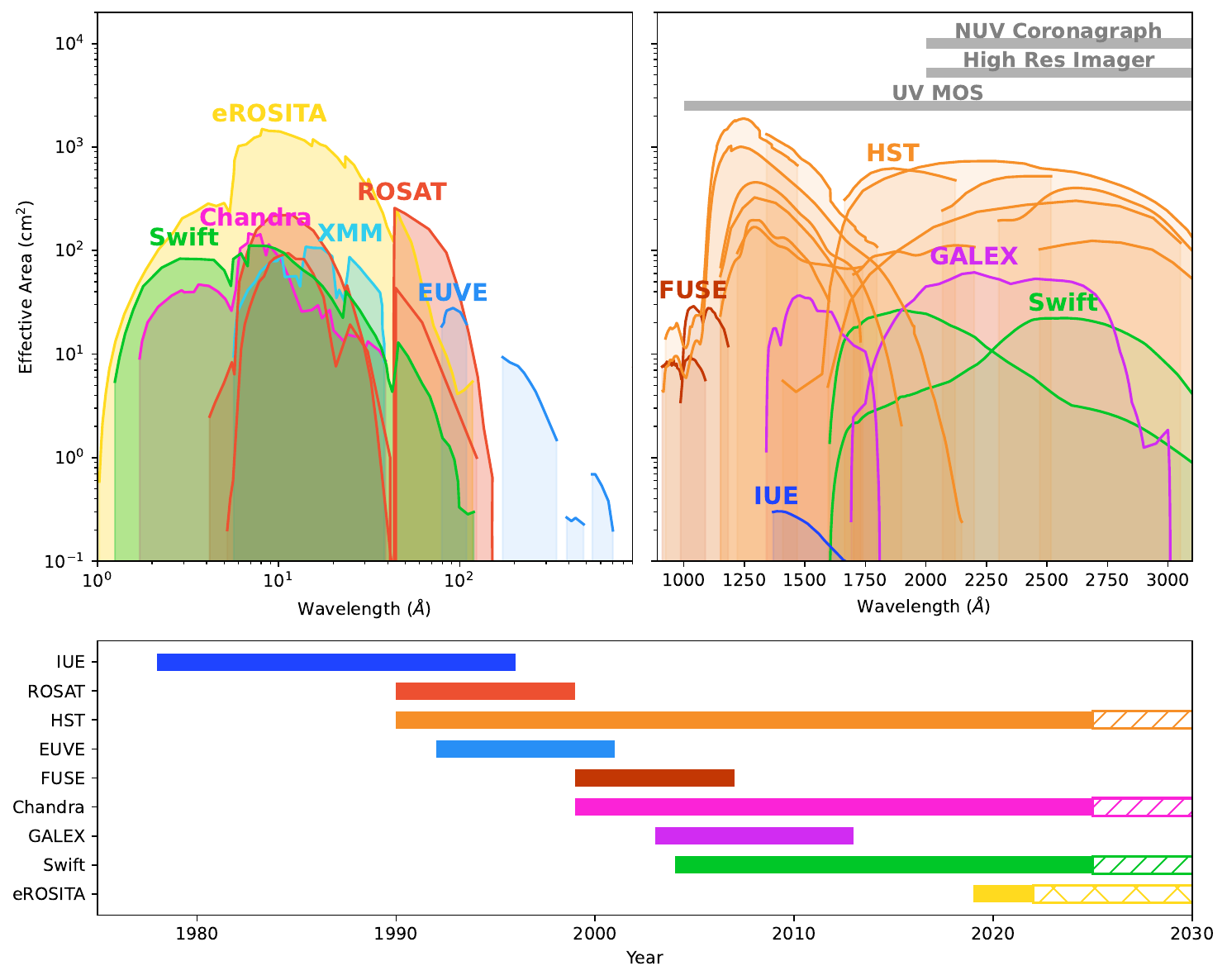}
    \caption{Effective area curves for all observatories used in this analysis. Gray bars mark the tentative wavelength coverage for HWO instruments: the UV Multi-Object Spectrograph (MOS; $\sim$1000--10\,000\,\AA), High Resolution Imager ($\sim$2000--25\,000\,\AA), and possible NUV coronagraph ($\sim$2000--4000\,\AA). \textit{Top left:} Effective area as a function of log wavelength for X-ray to EUV instruments. \textit{Top right:} Effective area in the UV, shown in linear wavelength space. (Only one effective area curve was available for IUE.) \textit{Bottom:} Operational lifetimes of each observatory, ordered by commissioning date. Solid bars indicate periods of active observation; hatched extensions denote projected future lifetimes. We note that eROSITA operations have been paused since early 2022. }
    \label{fig:missions}
\end{figure}

To evaluate the extent and limitations of existing high-energy data, it is important to consider the capabilities of both active and retired observatories. Space-based facilities such as HST, Galaxy Evolution Explorer (GALEX), and Far Ultraviolet Spectroscopic Explorer (FUSE) have provided extensive UV coverage, while X-ray observations have been obtained from Chandra, XMM-Newton, Swift, the R\"Oentgen SATellite (ROSAT), and more recently, the extended R\"Oentgen Survey with an Imaging Telescope Array (eROSITA). Many of these missions have already concluded operations or face uncertain futures, and none were designed with comprehensive coverage of all nearby HWO-relevant stars in mind. Their combined datasets, however, remain invaluable for constraining stellar activity and reconstructing panchromatic spectra. Figure \ref{fig:missions} summarizes the wavelength coverage, relative sensitivity, and operational lifetimes (past, present, and anticipated) of these key facilities. 

In this paper, we assess the current state of archival high-energy observations for the highest-priority HWO target stars. We focus on a sample of approximately 100 nearby stars, selected from the most promising candidates (``Tier 1") in the HWO Target Stars and Systems 2025 list (TSS25, described in Sect.~\ref{sec:sample}). We then survey the available high-energy data for these stars across a broad suite of space-based observatories. These include X-ray observatories including eROSITA, Chandra, XMM-Newton, ROSAT, and Swift, as well as ultraviolet missions: the Extreme-Ultraviolet Explorer (EUVE), FUSE, the International Ultraviolet Explorer (IUE), GALEX, and HST. Our goal is to quantify the existing coverage in the X-ray and UV regimes, characterize the spectral breadth and temporal depth of available observations, and identify critical gaps in the archival dataset. This analysis informs both the immediate needs for stellar characterization in support of HWO's science goals and the long-term planning of complementary analysis and observing campaigns.

\section{Sample Description}\label{sec:sample}

This analysis focuses on the highest priority target stars for HWO. As part of the \textit{Living Worlds} Science Working Group, the Target Stars and Systems (TSS) sub-working group constructed the TSS25 list of potential target stars for HWO's direct imaging survey, compiling the samples from existing catalogs of HWO targets \citep{Tuchow2025}. Among these existing catalogs were the HWO Preliminary Input Catalog \citep[HPIC,  ][]{Tuchow2024} and the NASA Exoplanet Exploration Program’s Mission Star List for HWO \citep[hereafter referred to as the
“ExEP list”, ][]{mamajek2023}. The TSS25 list contains approximately 13,000 stars, broken into three tiers of objects based on their likelihood of being observed by HWO and their potential contribution to the mission's science output. This study focuses on a subset of Tier 1 stars, which presently consists of the stars from the ExEP HWO target list \citep[][]{mamajek2023}\footnote{A queryable table of the ExEP HWO target stars can be found at the NASA Exoplanet Archive at \url{https://exoplanetarchive.ipac.caltech.edu/cgi-bin/TblView/nph-tblView?app=ExoTbls&config=DI_STARS_EXEP}. Additional stellar data for the ExEP HWO target stars were compiled by \citet{Harada2024}.}.

TSS25 Tier 1 comprises the best targets for a direct imaging survey for exoEarths. It consists of the 164 stars in the ExEP list that were selected based on the observability of Earth-sized planets in the habitable zone with a 6-meter class space telescope. 
Stars were selected on the basis of the accessibility of the habitable zone outside the coronagraph inner working angle and the estimated planet-star contrast. Most of these stars are main sequence FGK stars within 25 pc, along with a few of the closest and brightest M dwarfs.
The ExEP list separates three tiers of targets based on the presence of potential obstacles to direct imaging, such as the presence of circumstellar disks or binary companions. These tiers are labeled A, B, and C. Tier A consists of target stars where exoEarth candidates would be the brightest in reflected light, have the widest angular separation, and lack any other obstacles to direct imaging. Tiers B and C relax these criteria, allowing for dimmer exoEarth magnitudes (deeper expected contrasts) and smaller angular separation, and allow for disks or binary companions at a given brightness or separation. For this study, we restricted our sample to the highest-priority stars (ExEP Tiers A and B) resulting in a final list of 98 stars. This downselection from the full TSS25 Tier 1 list was necessary to keep the scope manageable with regard to validating the data across multiple observatories.

\section{Data Compilation Methods}\label{sec:data}

Because the structure, search capabilities, and data products of each observatory archive differ significantly, we adopted observatory-specific strategies for identifying and evaluating high-energy observations of the likely target stars. Each subsection below details the process used for a given mission, including how we queried the archive, selected relevant data, and interpreted observation metadata or spectra. While the methods varied across archives, we applied a consistent set of criteria for assessing whether a target was detected with usable data, detected with low quality or caveats, flagged as an upper limit, or affected by special cases such as unresolved binaries. This classification framework enables a coherent analysis of the breadth and quality of archival high-energy coverage across the stellar sample. Table \ref{tab:obs_summary} summarizes the number of observations available for each star from each observatory. Additional details and analysis of this dataset are provided in Section \ref{sec:obs_coverage}.

\subsection{eROSITA}\label{sec:eROSITA}
The eROSITA instrument aboard the Spectrum-Roentgen-Gamma (SRG) mission provides soft X-ray coverage in the 0.2–2.3~keV regime where stellar coronae emit most strongly
\citep{Merloni2012,Predehl2021,Merloni2024}. eROSITA's wide field of view (FOV) and all-sky coverage provide a uniform snapshot of soft X-ray emission from nearby stars; however, the relatively large positional uncertainties associated with survey-mode data (average astrometric uncertainty $\sim$10$\arcsec$–1$\arcsec$ in eRASS1; full-sky eROSITA first survey; \citealt{Merloni2024}, plus $<$10$\arcsec$ for strong detections) increase the risk of source confusion, especially in crowded fields or for unresolved binaries. Additionally, as a scanning mission designed for all-sky coverage, eROSITA's exposure time per source is relatively shallow, limiting sensitivity to only the brightest coronal sources within the stellar sample, though it is deeper than the ROSAT All-Sky Survey. The mission's early observation epochs (2019–2020) also restrict time-domain assessments, and optical follow-up is often needed to confirm stellar origin of faint detections or rule out contamination from background AGN. Furthermore, the public data release currently covers only half the sky, and observations have been paused since early 2022. Because eROSITA provides integrated count rates and broad-band flux estimates without resolving individual spectral lines, we classify its data as photometric rather than spectroscopic. 

To identify detections of target stars, a cross-match was performed between the stellar sample coordinates and the eRASS1 Main Catalog \citep{Merloni2024}\footnote{\url{https://vizier.cds.unistra.fr/viz-bin/VizieR-3?-source=J/A\%2bA/682/A34/erass1-m}}, which includes sources detected using the eSASS pipeline with likelihood values greater than 6. Proper motions were applied to the target coordinates prior to conducting a cone search with a 15$''$ radius, consistent with the average eROSITA astrometric uncertainty. This procedure yielded matches for a total of 37 target stars within the first eROSITA public data release, which covers the initial six months of the mission (December 2019–June 2020). While the proximity of many target stars makes them likely to be the dominant source within the eROSITA error circle, follow-up with higher-resolution facilities such as Chandra or XMM-Newton is ideal to confirm source identity, identify potential binary contributions, and rule out coincident extragalactic sources. As follow-up observations were beyond the scope of this work, we treated all detections as usable; the corresponding eROSITA data are provided in column 5 of Table \ref{tab:obs_summary}.

\subsection{Chandra}\label{sec:CHANDRA}

The Chandra X-ray Observatory is optimized for high-resolution imaging and spectroscopy in the soft X-ray regime (0.5--7~keV; $\sim$1--25~\AA), offering the best spatial resolution ($\sim$0.5$^{\prime\prime}$) of any X-ray telescope to date \citep{Weisskopf2002,Garmire2003}. Its sub-arcsecond spatial resolution makes Chandra uniquely capable of resolving individual stellar sources in crowded or binary systems. However, due to the limited FOV and long integration times required, its observations are biased toward well-studied, nearby, and often X-ray-bright stars.

Chandra data were compiled from the ChaSeR Chandra archive\footnote{\url{https://cda.harvard.edu/chaser/}}, which was queried for all targets and are contained in the Chandra Data Collection ~\dataset[DOI:10.25574/cdc.455]{https://doi.org/10.25574/cdc.455}. A total of eight stars were found to have been observed, although three were not detected. All observations were performed with the ACIS instrument, which covers the 0.5--7~keV energy range. Each detected star has been analyzed in detail by \citet{Binder2024}, including full crossmatching, spectral modeling with Astrophysical Plasma Emission Code (APEC) models, and uniform reprocessing and quality assessment of the imaging data. That work provides the most complete and consistent Chandra analysis to date for this sample, and therefore we have adopted their results directly. The Chandra data are provided in column 6 of Table \ref{tab:obs_summary}.

\subsection{XMM-Newton}\label{sec:XMM}

The XMM-Newton observatory covers a broad soft X-ray energy range (0.15--12~keV; $\sim$1--80~\AA) and is well-suited for detecting moderate-activity stars \citep{Jansen2001,Struder2001,Turner2001}. Its combination of high effective area and long exposures enables high-quality spectra even for moderately faint stellar coronae. XMM-Newton's wide FOV and more frequent survey-type programs introduce an observational bias toward serendipitously observed sources and stars in crowded fields.

We queried the XMM-Newton Science Archive\footnote{\url{https://nxsa.esac.esa.int/nxsa-web/\#search}} for all targets by their Henry Draper (HD) designation, which are cross-referenced by SIMBAD to identify coordinates and return matching observations. A total of 28 stars were retrieved, including two unresolved binaries. Of these, five targets were not detected. All observations were processed and analyzed by \citet{Binder2024}, who uniformly reduced the data and modeled each spectrum using APEC thermal plasma models across the 0.8--62~\AA\ range. For undetected sources, fluxes were estimated based on sensitivity limits; we have adopted their results. XMM-Newton observational data are compiled in column 7 of Table \ref{tab:obs_summary}.

\subsection{ROSAT}\label{sec:ROSAT}

ROSAT conducted the first imaging all-sky survey in soft X-rays, operating in the 0.1–2.4~keV energy range with the Position Sensitive Proportional Counter (PSPC) \citep{Truemper1982,Voges1999,Voges2000,Boller2016}. However, the moderate spatial resolution ($\sim$25 arcsec) and sensitivity of the mission introduce important limitations. Source confusion is possible in crowded fields, and unresolved binaries or faint background sources may bias flux estimates. Additionally, the shallow exposure of the all-sky survey means that only the most X-ray luminous stars were detected, strongly biasing the catalog toward younger, more active stars \citep[e.g.][]{Freund2022}. Because ROSAT provided broadband fluxes without spectral resolution sufficient to isolate individual lines or detailed continuum features, we classify its data as photometric.

We cross-matched the HD names of the target stars with the second ROSAT all-sky survey (2RXS) catalog \citep{Boller2016}. This catalog is an extended and revised version of their first publicly released ROSAT catalog, the All-Sky Bright Source Catalog and Faint Source Catalog \citep[1RXS;][]{Voges1999,Voges2000}. Of the target stars, a total of 70 have reported count rates in 2RXS. Four of the 70 stars do not have a 1RXS counterpart, and an additional three lack reported hardness ratios. We flag these seven stars as having potential issues. The 2RXS catalog does not separate binary star names by component, so we also flag 11 stars that are likely unresolved binaries. We assume the other 52 stars have reliable ROSAT X-ray detections, but conduct no further data quality analysis. Column 8 of Table \ref{tab:obs_summary} lists the ROSAT observations.

\subsection{EUVE}\label{sec:EUVE}

EUVE operated from 1992 to 2001 and provided photometric and spectroscopic coverage across several bands in the EUV (70--760\,\AA) \citep{Bowyer1991,Bowyer1994,Bowyer1996}, a wavelength regime that is critical for characterizing high-energy stellar emission and estimating atmospheric mass loss from exoplanets. However, there are several important observational biases and limitations associated with the EUVE dataset. First, EUV photons are strongly absorbed by interstellar neutral hydrogen and helium, which severely limits detections to very nearby stars-typically within tens of parsecs \citep[e.g.][]{Craig1997}. Second, most of the detected stars were not observed with EUVE's spectrometers, leaving only broadband photometry that provides limited diagnostic power for line-dominated EUV spectra. Third, the mission's sensitivity favored active solar-type, late-type stars and white dwarfs, but not solar-type stars with moderate activity levels. Finally, EUVE was designed and operated before the discovery of transiting exoplanets and before the importance of EUV radiation for planetary atmosphere evolution was widely appreciated. As a result, the mission's observing strategy was not optimized to support exoplanet science, and most of its stellar targets were selected for unrelated science goals.

We used SIMBAD to cross-reference the target list with the EUVE Deep Survey point source catalog \citep{Bowyer1996}, which includes EUVE-specific identifiers. We identified 24 stars with a photometric detection in at least one EUVE band, typically in the 70–190~\AA\ range, where the instrument had its highest sensitivity and where interstellar attenuation is lowest. We then compared these detections with the catalog of EUVE spectroscopic observations. Only two targets in our sample, $\kappa^1$~Cet (HD 20630) and $\xi$~Boo~A (HD 131156A), had spectra that appeared qualitatively distinct from the sky background. While an additional high-quality EUVE spectrum exists for $\alpha$~Cen~A+B (HD 128620 and HD 128621), the binary is unresolved, so we flag these stars as having data with known issues. The summary of EUVE data is given in column 9 of Table \ref{tab:obs_summary}, and the corresponding EUVE observations can be accessed via \dataset[doi:10.17909/cs47-4y74]{https://doi.org/10.17909/cs47-4y74}.

\subsection{Swift}\label{sec:Swift}

The Neil Gehrels Swift Observatory conducts rapid-response, short-exposure X-ray observations (0.3--10~keV) using the X-Ray Telescope (XRT) instrument \citep{Gehrels2024,Burrows2005}. Due to its short integration times (typically $\lesssim 2$~ks), Swift is most effective for capturing transient events or providing upper limits for bright, nearby stars. Swift’s flexible scheduling allows for rapid-response or snapshot observations that can complement longer campaigns; however, this also introduces a bias towards more active and/or previously known sources. Its Ultraviolet/Optical Telescope (UVOT) adds the ability to detect NUV fluxes in a similar bandpass to GALEX and offered simultaneous UV and X-ray coverage, allowing for time-resolved studies of stellar flares and variability \citep{Roming2005}. Because Swift's XRT provides broadband fluxes and count rates without resolving individual emission lines, we classify the data as photometric rather than spectroscopic.

Targets were cross-referenced by HD name with the Swift archive using NASA’s HEASARC Swift search portal\footnote{\url{https://heasarc.gsfc.nasa.gov/cgi-bin/W3Browse/swift.pl}}. A total of 21 Tier 1 stars were imaged with the XRT instrument, of which four are unresolved binaries. Seven of the 21 stars were also detected with the UVOT/UM2 filter (1600--3110\,\AA). Most XRT exposures resulted in flux upper limits, consistent with the short integration times; this data quality vetting was performed by \citet{Binder2024}. The Swift data summary is provided in column 10 of Table \ref{tab:obs_summary}.

\subsection{FUSE}\label{sec:FUSE}
FUSE observed the FUV bandpass from 912--1187 \AA\ between 1999 and 2007 \citep{Moos2000,Sahnow2000}. This spectral window contains key diagnostics of stellar chromospheres, transition regions, and coronae, sampling plasma temperatures from $10^4$\,K to over $10^7$\,K. Notably, the FUSE band includes strong emission lines such as the hydrogen Lyman series and the O~\textsc{vi} doublet and LISM absorption in C~\textsc{iii} and N~\textsc{II}. However, the absence of onboard background subtraction and its low-Earth orbit led to frequent contamination from geocoronal emission and scattered solar radiation (particularly in the H~\textsc{I} and O~\textsc{I} lines) requiring careful interpretation \citep{Feldman2001, Dixon2007}. Additionally, FUSE's stellar target list was driven by specific science goals, resulting in a sample skewed toward younger or more active stars and not necessarily representative of the HWO target population.

We used MAST's FUSE search portal\footnote{\url{https://archive.stsci.edu/fuse/}} to cross-reference the target list with all available FUSE spectral observations, searching by HD name and a search radius of 3\,arcmin. For high proper motion stars, the search coordinates were propagated to the appropriate epoch of FUSE observations to ensure accurate cross-matching. Only eight targets were observed with FUSE, consistent with the mission's focus on relatively bright or high-priority UV sources (see, e.g., \citealt{Redfield2002} for a FUSE coronal activity survey of solar-like stars). For each matching observation, we compiled the associated program identifier and verified the presence of clearly detected FUV spectra. We also examined the Science Data Assessment Forms for each of the observations to identify any issues, of which only one observation was flagged. We consider the remaining observations to be of usable quality, though we did not conduct a detailed emission line analysis in this work and a future step should involve careful identification of stellar versus background features to determine the scientific utility of each spectrum. The FUSE data is summarized in column 11 of Table~\ref{tab:obs_summary}, and the corresponding FUSE observations can be accessed via \dataset[doi:10.17909/8vcq-4175]{https://doi.org/10.17909/8vcq-4175}.

\subsection{IUE}\label{sec:iue}
IUE provided UV spectra in both short-wavelength (SWP; 1150–2000\,\AA) and long-wavelength (LWP/LWR; 1850–3350\,\AA) modes, with both low- and high-resolution settings \citep{Boggess1978,Ayres1981}. While IUE represents one of the earliest (1978--1996) large UV spectral archives and was foundational for early stellar atmosphere modeling, its data are known to suffer from serious calibration limitations. In particular, early assessments revealed significant scattered light contamination from the NUV into the FUV channel, as well as flux calibration inconsistencies that may compromise scientific use of the low-resolution spectra, especially for faint late-type stars \citep{Basri1985}. The quality of observed LWP/LWR spectra degrades bluewards of $\lesssim 2300$\,\AA\ for late-type stars \citep{Chavez2007} and \lya\ emission in the SWP mode is also dominated by geocoronal airglow emission \citep{France2016} that obscures other emission lines such as N~\textsc{v} and Si~\textsc{iii} \citep{Kamgar2024}. These issues must be taken into account when evaluating the utility of IUE data for characterizing stellar UV environments.

We retrieved a list of available IUE datasets for our targets from the MAST archive\footnote{\url{https://archive.stsci.edu/iue/}}, using HD numbers from the target list and a search radius of 3\,arcmin. For high proper motion stars, coordinates were propagated to the epoch of the IUE observation to ensure accurate cross-matching with archival datasets. Because the archive does not reliably differentiate between components of binary systems sharing the same HD number, confusion can arise in identifying which star was observed. For example, HD 165341 B was removed from the final list after visual inspection confirmed that only HD~165341 A was observed.

We classified the data quality of the observations using parameters from NEWSIPS-processed data\footnote{The New Spectral Image Processing System (NEWSIPS) is the second, and final, generation standard production processing system used for IUE data.}, assigning each as usable, having ``known issues'', or ``poor''. An observation was classified as usable only if all relevant parameters exceeded specific thresholds: a ratio of maximum continuum count to background (a quick signal-to-noise ratio estimate) greater than 5, a cross-correlation success rate (i.e., the percentage of spectral regions that successfully cross-correlate with the template) above 90\%, and a median cross-correlation coefficient (used to assess the reliability of the wavelength solution) above 0.85. If any of these parameters fell into intermediate ranges (signal-to-noise ratio between 2 and 5, cross-correlation success between 80\% and 90\%, or a median coefficient between 0.70 and 0.85) the observation was flagged as having ``known issues''. Values below these thresholds resulted in a ``poor'' classification. However, if a high-level science product 
was available for an observation, it was automatically classified as usable regardless of the parameter values.

We note that this inspection does not fully address the scattered light contamination described above, which remains a fundamental concern for interpreting IUE spectra in the FUV. And further issues with NEWSIPS, including flux calibration being inconsistent by nearly 10\% \citep{Massa2000}, means that if the data is not adequately corrected, these effects could render many of the available spectra scientifically unusable. Despite these limitations, we found that 62 of the 98 stars have at least some IUE data available, with 33 of these stars having data in at least one mode classified as usable. Column 12 of Table~\ref{tab:obs_summary} contains the summary of IUE data, and the corresponding IUE observations can be accessed via \dataset[doi:10.17909/fb84-tm31]{https://doi.org/10.17909/fb84-tm31}.

\subsection{GALEX}\label{sec:GALEX}

GALEX provided uniform, wide-area photometric coverage in both the FUV (1350--1750\,\AA) and NUV (1750--2800\,\AA), enabling statistical studies of stellar activity across large samples \citep{Morrissey2007,Bianchi2011,Bianchi2017}. Observations were conducted between 2003 and 2012, with sensitivity and dynamic range sufficient to detect many nearby stars relevant to exoplanet host characterization. However, the GALEX detectors were subject to significant deadtime corrections for the brightest stars and suffered from significant photometric uncertainties in the low signal-to-noise regime, particularly in the FUV. In addition, the spatial resolution of $\sim$5\arcsec\ could lead to source confusion in crowded fields.

To identify GALEX detections among the target stars, we corrected the stellar coordinates to epoch 2007 (the median epoch for GALEX observations) using proper motions from Gaia DR3 \cite{Gaia2021}. We then queried the GALEX MAST catalog\footnote{\url{https://galex.stsci.edu/GR6/?page=mastform}} using \texttt{astroquery} \citep{Ginsburg2019}, searching within 6\arcsec\ of the expected positions. We excluded matches flagged for known artifacts, such as reflections or detector halos, and visually confirmed each detection by comparing to the GALEX tile images to ensure the source aligned with the expected stellar position. This process identified 36 of the 98 stars with viable photometric measurements in at least one band. Of these, 34 stars had FUV data and 13 had NUV data, with 11 stars detected in both bands. Ten of the 13 NUV detections fall within the non-linear regime of the GALEX NUV detector (magnitude $NUV <$ 15\,mag), as characterized by \citet{Camarota2014}, and can be corrected using the empirical relations from \citet{Wall2019}. These are flagged as ``having known issues''. In addition to these detections, 21 stars were observed but not detected in the NUV, and 4 were similarly undetected in the FUV. While these nondetections could be treated as upper limits, we instead classify them as ``failed'' observations and exclude them from the GALEX summary presented in column 13 of Table~\ref{tab:obs_summary}. The corresponding GALEX observations can be accessed via \dataset[doi:10.17909/86jf-jp84]{https://doi.org/10.17909/86jf-jp84}.

\subsection{HST}\label{sec:hst}

HST provides low-, medium- and high-resolution ultraviolet spectroscopy across both the far-ultraviolet (FUV; $\sim$1060--1750~\AA) and near-ultraviolet (NUV; $\sim$1750--3200~\AA) regimes through multiple observing modes, primarily using the Cosmic Origins Spectrograph (COS; \citealt{Green2012}) and the Space Telescope Imaging Spectrograph (STIS; \citealt{Kimble1998}). Each instrument offers several grating and detector combinations, enabling flexibility in spectral resolution ($\mathcal{R}~\sim$1,000--100,000) and wavelength coverage tailored to specific science goals. This versatility makes HST ideal for probing chromospheric, transition region, and coronal diagnostics relevant to exoplanet host star characterization. While individual HST programs typically target narrow wavelength windows, two curated archival products -- the Low Resolution Stellar Library (LOWLIB) and the Hubble Advanced Spectral Products (HASP) -- aggregate heterogeneous spectroscopic datasets into standardized, accessible formats for community use. Other archival products of nearby stars are also available, but not used further in this study (e.g., ASTRAL \citep{Ayres2014} and StarCAT \citep{Ayres2010})

Despite its strengths, HST UV spectroscopy does have observational biases. Its narrow FOV and pointed, program-specific observations limit the number of stars with complete UV coverage, particularly among fainter or less active stars. Bright target limits for COS and STIS can also preclude observations of the most nearby or UV-luminous stars without special observing modes, introducing sample incompleteness. Scheduling constraints and limited long-term availability of UV modes have further restricted the total volume of stellar UV data acquired by HST. Moreover, neither instrument provides continuous high-resolution coverage across the full FUV or NUV bandpasses, limiting access to certain spectral regions and requiring trade-offs in grating and mode selection.

LOWLIB\footnote{\url{https://archive.stsci.edu/hlsp/lowlib}} compiles STIS CCD spectra from a range of programs, covering 1,710--10,070\,\AA, and applies enhanced processing to improve the quality of the CCD-mode data. HASP\footnote{\url{https://archive.stsci.edu/missions-and-data/hst/hasp}}, in contrast, coadds and splices all available COS and STIS spectra for each target, providing more comprehensive wavelength coverage but with less optimized handling of CCD-mode observations. Because of these complementary strengths, we include results from both catalogs in our analysis. To assess HST coverage for our targets, we crossmatched our sample with both datasets. For LOWLIB, we matched directly to the published target list and identified 27 stars with available spectra. For HASP, we queried the HST archive using each star's HD name and retrieved the corresponding HASP products. For each match, we compiled the instrument modes (i.e., detector and grating combinations) and classified the observations as FUV or NUV. In total, we found that 25 stars have data in at least one FUV mode, and 30 have data in at least one NUV mode. Twenty-one targets have coverage in both regimes.

We further crossmatched our targets with the MAST archive\footnote{\url{https://mast.stsci.edu/search/ui/\#/hst}} using HD names and a search radius of 3\,arcmin.  We retrieved all COS and STIS Science Spectrum results with central wavelengths below 3000\,{\AA}. After manually removing any white dwarfs from the list, we identified 10 additional targets not included in the LOWLIB or HASP counts. We visually inspected these spectra and confirmed they are of usable quality. Of these 10, 3 have FUV-only data, 1 has NUV-only, and 6 have both FUV and NUV.

Earlier in its mission, HST also employed the Goddard High Resolution Spectrograph (GHRS), which operated from 1990 to 1997 and provided high-resolution spectroscopy in the 1150--3200\,\AA\ range. GHRS was particularly well-suited to detailed studies of individual UV lines, including \lya, Mg~\textsc{II}, and C~\textsc{IV}, and contributed several benchmark datasets for solar analogs and active stars. However, GHRS datasets are less uniformly calibrated than more recent STIS and COS spectra, and fewer targets were observed due to the limited early-mission UV focus and narrower science drivers before the exoplanet era. We crossmatched our sample with GHRS data in MAST (using the same 3\,arcmin radius and HD designations) and retrieved all relevant ``Science Spectrum" results. Thirteen targets were observed with GHRS, but only 11 have usable observations (based on data quality flags reported in MAST's quick view). Of these 11, 1 has FUV-only data, 6 have NUV-only, and 4 have both FUV and NUV.

In total, 52 stars in our sample have usable-quality COS, STIS, and/or GHRS ultraviolet spectra, with 4 showing FUV-only coverage, 18 with NUV-only, and 30 with data in both regimes. The HST data summary is presented in the final column of Table~\ref{tab:obs_summary}, and the corresponding HST observations can be accessed via \dataset[doi:10.17909/kvmm-j785]{https://doi.org/10.17909/kvmm-j7854}.

\section{Observational Coverage Analysis} \label{sec:obs_coverage}

This section quantifies the availability and quality of multiwavelength data for each target, organized by observatory, wavelength band, and data type. A catalog of the information is available in machine-readable format online. Table~\ref{tab:columns} describes the columns in the catalog.

\begin{deluxetable*}{lcc}[htb]
\tablenum{1}
\tablecaption{Description of columns in the HWO TSS High Energy Emission data catalog}
\tablehead{Column Number & Column Name & Description}
\label{tab:columns}                
\startdata
    1 & hd$\_$name &  Henry Draper Catalog ID \\
    2 & tic$\_$name &  TESS Input Catalog ID \\
    3 & ra	 & Right ascension at epoch J2000 (ICRS) \\
    4 & dec	 & Declination at epoch J2000 (ICRS) \\
    5 & v$\_$mag & Johnson $V$ magnitude \\
    6 & dist &  Distance (pc) \\
    7 & spt	 & Spectral type \\
    8 & teff & Effective temperature (K) \\
    9 & exep$\_$tier & ExEP tier classification (A/B) \\	
    10 & observatory &  Observatory used for the observation\\
    11 & instrument	& Instrument used for the observation\\
    12 & mode & Observing mode \\
    13 & type & Observation type (spectroscopy/photometry) \\
    14 & band & General wavelength band (X-ray/EUV/FUV/NUV) \\
    15 & wl$\_$coverage  & Wavelength coverage of observation (\AA) \\
    16 & data$\_$quality & Description of data quality \\
    17 & quality$\_$flag & \makecell[c]{0 = poor quality data\\1 = usable data\\2 = data with known issues}\\
    18 & url & URL for where to access the data\\
\enddata
\tablecomments{This table is available in its entirety in machine-readable form. Columns 2--8 come directly from the HPIC catalog \citep{Tuchow2024} and the references within.} 
\end{deluxetable*}

\begin{figure}
    \centering
    \includegraphics[width=0.95\linewidth]{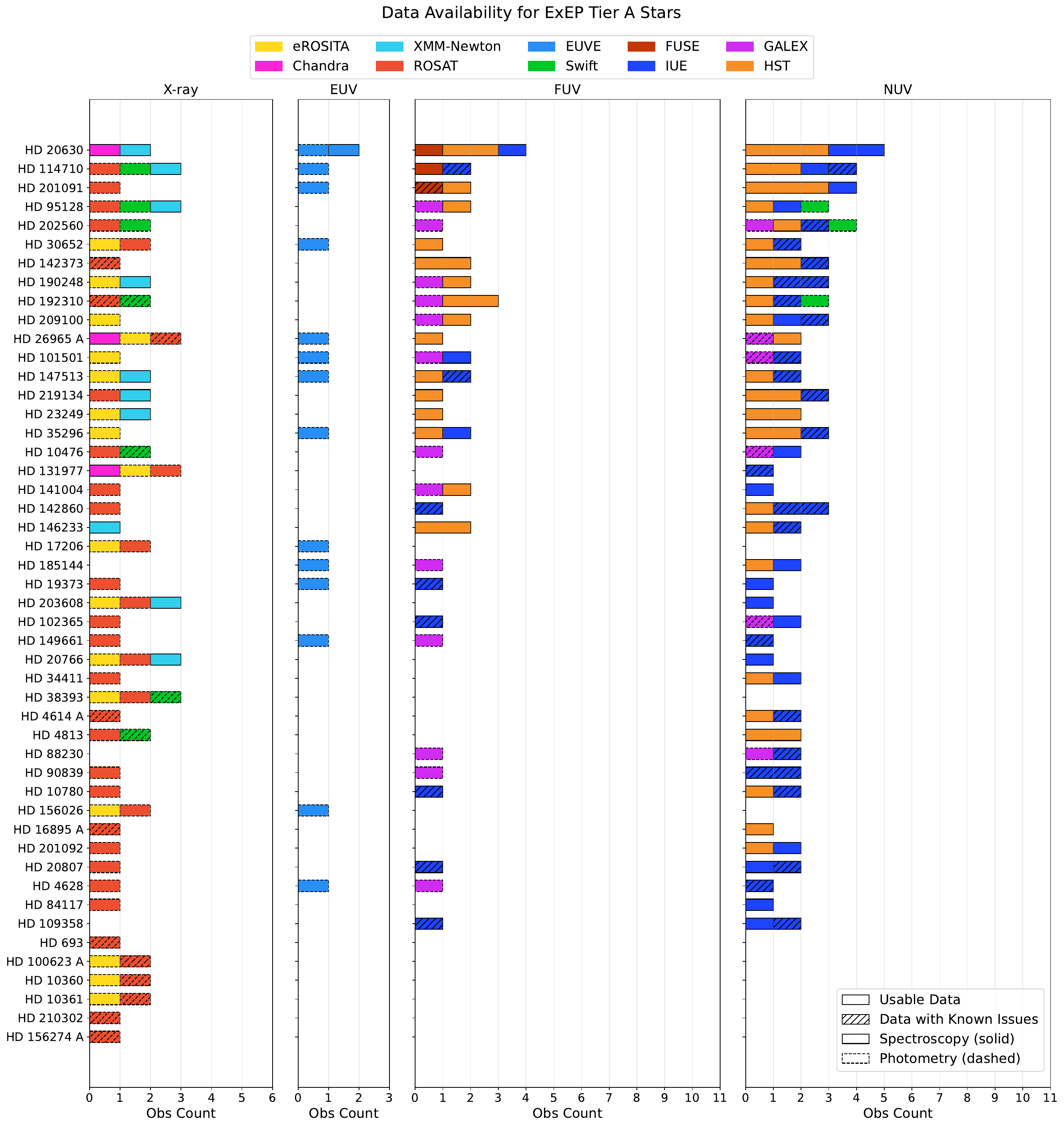}
    \caption{Data availability across X-ray, EUV, FUV, and NUV wavelength ranges for ExEP Tier A stars, ordered by total number of observations (highest at top). Each colored block represents an individual observation, color-coded by observatory. Solid blocks indicate usable data, while hatched blocks denote flagged data with known issues. These include observations of unresolved binaries (Swift, XMM-Newton), upper limits (Swift), or data requiring linear corrections (GALEX). Blocks with solid outlines represent spectroscopic measurements (including APEC model spectra for Chandra and XMM-Newton observations), while dashed outlines represent photometry. Subplots reflect different wavelength bands.}
    \label{fig:exepA_data}
\end{figure}

\begin{figure}
    \centering
    \includegraphics[width=0.95\linewidth]{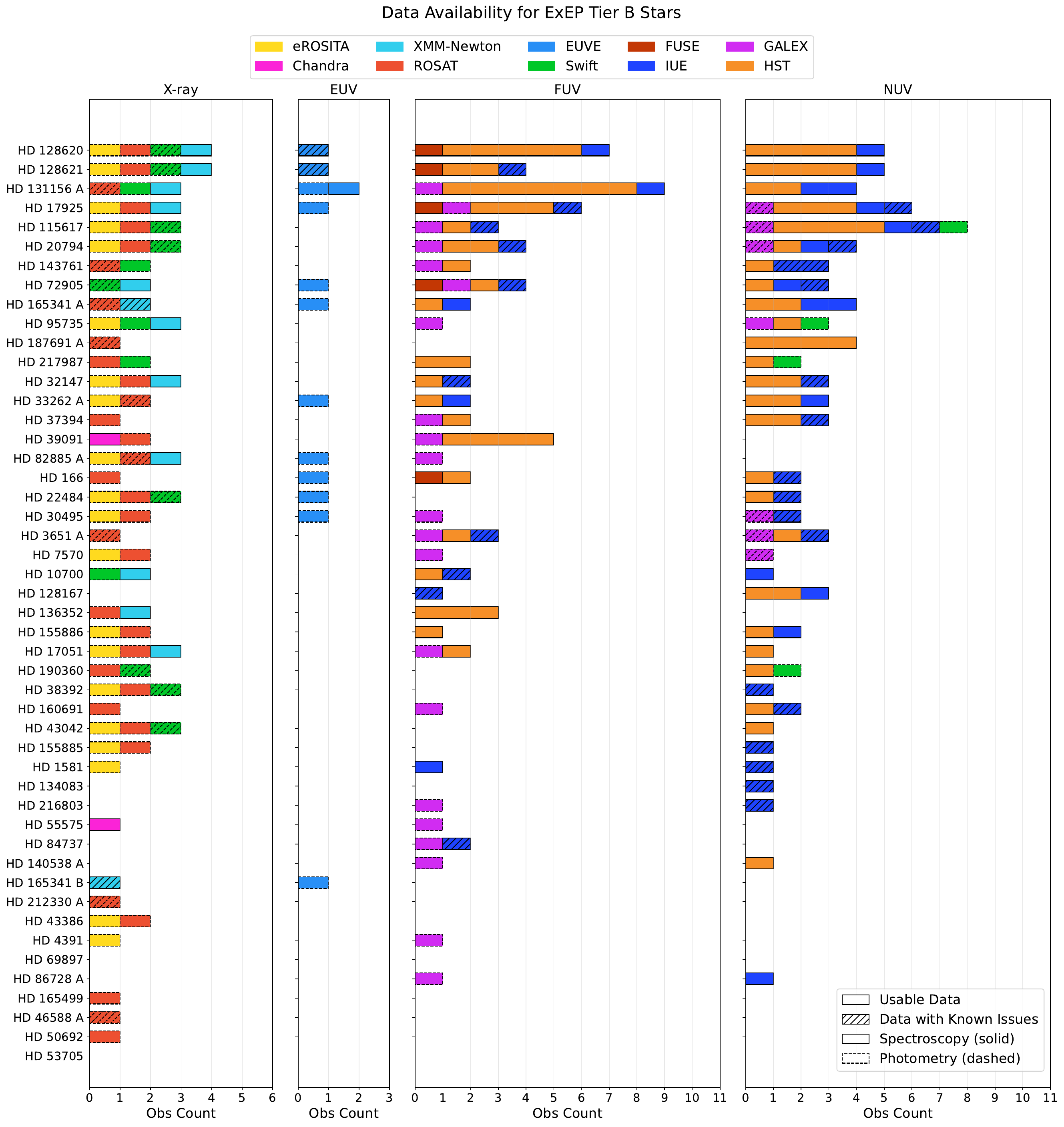}
    \caption{Same as Figure \ref{fig:exepA_data}, but for ExEP Tier B stars.}
    \label{fig:exepB_data}
\end{figure}

Table~\ref{tab:obs_summary} provides a comprehensive overview of the number of observations available for each star as recorded by different observatories. Johnson $V$ magnitudes span 0.01--7.52\,mag, distances range from 1.35 to 21.78,pc, and effective temperatures ($T_{\rm eff}$) from 3469 to 6782\,K, with values drawn from the HPIC catalog \citep{Tuchow2024} and references therein. For each star–observatory pair, the table lists the number of distinct observing visits and indicates data quality according to the criteria defined in Sect.~\ref{sec:data}. To visualize this information, Fig. \ref{fig:exepA_data} and Fig. \ref{fig:exepB_data} show stacked bar plots for ExEP Tier A and Tier B stars, respectively, organized by total number of usable-quality observations. The sample was divided into two figures both for clarity (given the large number of targets) and to reflect the ExEP-defined prioritization, with Tier A comprising the higher-priority candidates for future observations. These figures highlight wavelength-specific coverage, making it easy to identify stars with panchromatic data or significant gaps. We note that some stars have counts $>$1 for specific observatories in both Table \ref{tab:obs_summary} and Figures \ref{fig:exepA_data} and \ref{fig:exepB_data}. In most instances, this number is reflective of observations being taken in various observing modes, however, a small fraction of stars do have repeat measurements in a specific mode that allow for variability analysis (predominantly with HST data).


\begin{ThreePartTable}
\begin{TableNotes}
\item \textbf{Note.} \strut Observatory counts are shown as follows: plain numbers indicate usable data; numbers in parentheses denote data with known issues (e.g., unresolved binaries, upper limits); numbers in  italics and square brackets indicate poor-quality data.
\end{TableNotes}

\begin{longtable}{p{1.9cm} c c c *{10}{c}}

\caption{Observation summary by observatory.}\label{tab:obs_summary}\\
\toprule
Star Name & $V$ (mag) & $d$ (pc) & $T_{\rm eff}$ (K) & \rot{eROSITA} & \rot{Chandra} & \rot{XMM} & \rot{ROSAT} & \rot{EUVE} & \rot{Swift} & \rot{FUSE} & \rot{IUE} & \rot{GALEX} & \rot{HST} \\
\hline
\endfirsthead
\multicolumn{14}{c}{{\bfseries \tablename\ \thetable{} -- continued from previous page}} \\
\toprule
Star Name & Vmag & Dist & Teff & \rot{eROSITA} & \rot{Chandra} & \rot{XMM} & \rot{ROSAT} & \rot{EUVE} & \rot{Swift} & \rot{FUSE} & \rot{IUE} & \rot{GALEX} & \rot{HST} \\
\hline
\endhead
\hline 
\multicolumn{14}{r}{{Continued on next page}} \\
\insertTableNotes
\endfoot
\hline
\endlastfoot
HD 100623 A & 5.98 & 9.56 & 5196 & 1 & \nodata & \nodata & (1) & \nodata & \nodata & \nodata & \nodata & \nodata & \nodata \\
HD 101501 & 5.34 & 9.57 & 5491 & 1 & \nodata & \nodata & \nodata & 1 & \nodata & \nodata & 1, (1), \textit{[1]} & 1, (1) & \nodata \\
HD 102365 & 4.88 & 9.32 & 5618 & \nodata & \nodata & \nodata & 1 & \nodata & \nodata & \nodata & 1, (1), \textit{[1]} & (1) & \nodata \\
HD 10360 & 5.68 & 8.19 & 5025 & 1 & \nodata & \nodata & (1) & \nodata & \nodata & \nodata & \nodata & \nodata & \nodata \\
HD 10361 & 5.80 & 8.20 & 5111 & 1 & \nodata & \nodata & (1) & \nodata & \nodata & \nodata & \nodata & \nodata & \nodata \\
HD 10476 & 5.24 & 7.64 & 5204 & \nodata & \nodata & \nodata & 1 & \nodata & (1) & \nodata & 1, \textit{[1]} & 1, (1) & \nodata \\
HD 10700 & 3.50 & 3.65 & 5356 & \nodata & \nodata & 1 & \nodata & \nodata & 1 & \nodata & 1, (1), \textit{[1]} & \nodata & 1 \\
HD 10780 & 5.63 & 10.04 & 5358 & \nodata & \nodata & \nodata & 1 & \nodata & \nodata & \nodata & (2) & \nodata & 1 \\
HD 109358 & 4.25 & 8.47 & 5878 & \nodata & \nodata & \nodata & \nodata & \nodata & \nodata & \nodata & 1, (2) & \nodata & \nodata \\
HD 114710 & 4.25 & 9.20 & 5996 & \nodata & \nodata & 1 & 1 & 1 & 1 & 1 & 1, (2), \textit{[1]} & \nodata & 2 \\
HD 115617 & 4.74 & 8.53 & 5552 & 1 & \nodata & \nodata & 1 & \nodata & 1, (1) & \nodata & 1, (2) & 1, (1) & 5 \\
HD 128167 & 4.47 & 15.75 & 6782 & \nodata & \nodata & \nodata & \nodata & \nodata & \nodata & \nodata & 1, (1), \textit{[1]} & \nodata & 2, \textit{[1]} \\
HD 128620 & 0.01 & 1.35 & 5776 & 1 & \nodata & 1 & 1 & (1) & (1) & 1 & 2, \textit{[1]} & \nodata & 9, \textit{[1]} \\
HD 128621 & 1.33 & 1.35 & 5244 & 1 & \nodata & 1 & 1 & (1) & (1) & 1 & 1, (1), \textit{[1]} & \nodata & 6, \textit{[4]} \\
HD 131156 A & 4.59 & 6.75 & 5620 & \nodata & \nodata & 1 & (1) & 2 & 1 & \nodata & 3, \textit{[1]} & 1 & 9 \\
HD 131977 & 5.72 & 5.89 & 4632 & 1 & 1 & \nodata & 1 & \nodata & \nodata & \nodata & (1), \textit{[2]} & \nodata & \nodata \\
HD 134083 & 4.93 & 19.53 & 6528 & \nodata & \nodata & \nodata & \nodata & \nodata & \nodata & \nodata & (1), \textit{[2]} & \nodata & \nodata \\
HD 136352 & 5.65 & 14.74 & 5685 & \nodata & \nodata & 1 & 1 & \nodata & \nodata & \nodata & \textit{[1]} & \nodata & 3 \\
HD 140538 A & 5.86 & 14.78 & 5682 & \nodata & \nodata & \nodata & \nodata & \nodata & \nodata & \nodata & \nodata & 1 & 1 \\
HD 141004 & 4.42 & 11.90 & 5898 & \nodata & \nodata & \nodata & 1 & \nodata & \nodata & \nodata & 1, \textit{[2]} & 1 & 1 \\
HD 142373 & 4.62 & 15.90 & 5820 & \nodata & \nodata & \nodata & (1) & \nodata & \nodata & \nodata & (1), \textit{[2]} & \nodata & 4 \\
HD 142860 & 3.84 & 11.16 & 6285 & \nodata & \nodata & \nodata & 1 & \nodata & \nodata & \nodata & (3), \textit{[1]} & \nodata & 1 \\
HD 143761 & 5.41 & 17.50 & 5812 & \nodata & \textit{[1]} & \nodata & (1) & \nodata & 1 & \nodata & (2), \textit{[1]} & 1 & 2, \textit{[1]} \\
HD 146233 & 5.50 & 14.13 & 5785 & \nodata & \nodata & 1 & \nodata & \nodata & \nodata & \nodata & (1), \textit{[1]} & \nodata & 3 \\
HD 147513 & 5.38 & 12.89 & 5868 & 1 & \nodata & 1 & \nodata & 1 & \nodata & \nodata & (2) & \nodata & 2 \\
HD 149661 & 5.77 & 9.89 & 5262 & \nodata & \nodata & \nodata & 1 & 1 & \nodata & \nodata & (1), \textit{[1]} & 1 & \nodata \\
HD 155885 & 5.03 & 5.95 & 5144 & 1 & \nodata & \nodata & 1 & \nodata & \nodata & \nodata & (1), \textit{[1]} & \nodata & \nodata \\
HD 155886 & 4.33 & 5.95 & 5132 & 1 & \nodata & \nodata & 1 & \nodata & \nodata & \nodata & 1, \textit{[1]} & \nodata & 2 \\
HD 156026 & 6.34 & 5.95 & 4476 & 1 & \nodata & \nodata & 1 & 1 & \nodata & \nodata & \textit{[1]} & \nodata & \nodata \\
HD 156274 A & 5.48 & 8.79 & 5235 & \nodata & \nodata & \nodata & (1) & \nodata & \nodata & \nodata & \nodata & \nodata & \nodata \\
HD 1581 & 4.23 & 8.61 & 5932 & 1 & \nodata & \nodata & \nodata & \nodata & \nodata & \nodata & 1, (1), \textit{[1]} & \nodata & \nodata \\
HD 160691 & 5.15 & 15.60 & 5761 & \nodata & \nodata & \textit{[1]} & 1 & \nodata & \nodata & \nodata & (1) & 1 & 1 \\
HD 165341 A & 4.21 & 5.11 & 5298 & \nodata & \nodata & (1) & (1) & 1 & \nodata & \nodata & 3 & \nodata & 3 \\
HD 165341 B & 6.07 & 5.10 & 4348 & \nodata & \nodata & (1) & \nodata & 1 & \nodata & \nodata & \nodata & \nodata & \nodata \\
HD 166 & 6.07 & 13.76 & 5491 & \nodata & \nodata & \nodata & 1 & 1 & \nodata & 1 & (1), \textit{[1]} & \nodata & 2 \\
HD 16895 A & 4.11 & 11.14 & 6263 & \nodata & \nodata & \nodata & (1) & \nodata & \nodata & \nodata & \textit{[2]} & \nodata & 1 \\
HD 17051 & 5.40 & 17.36 & 6157 & 1 & \nodata & 1 & 1 & \nodata & \nodata & \nodata & \nodata & 1 & 2 \\
HD 17206 & 4.46 & 14.27 & 6330 & 1 & \nodata & \nodata & 1 & 1 & \nodata & \nodata & \textit{[3]} & \nodata & \nodata \\
HD 17925 & 6.05 & 10.36 & 5199 & 1 & \nodata & 1 & 1 & 1 & \nodata & 1 & 1, (2) & 1, (1) & 6 \\
HD 185144 & 4.68 & 5.76 & 5298 & \nodata & \nodata & \nodata & \nodata & 1 & \nodata & \nodata & 1, \textit{[2]} & 1 & 1 \\
HD 187691 A & 5.12 & 19.48 & 6134 & \nodata & \nodata & \nodata & (1) & \nodata & \nodata & \nodata & \textit{[3]} & \nodata & 4 \\
HD 190248 & 3.56 & 6.10 & 5576 & 1 & \nodata & 1 & \nodata & \nodata & \nodata & \nodata & (2), \textit{[1]} & 1 & 2 \\
HD 190360 & 5.74 & 15.99 & 5563 & \nodata & \nodata & \textit{[1]} & 1 & \nodata & 1, (1) & \nodata & \textit{[1]} & \nodata & 1 \\
HD 192310 & 5.72 & 8.81 & 5087 & \nodata & \nodata & \nodata & (1) & \nodata & 1, (1) & \nodata & (1) & 1 & 3 \\
HD 19373 & 4.06 & 10.57 & 5952 & \nodata & \textit{[1]} & \nodata & 1 & 1 & \nodata & \nodata & 1, (1), \textit{[1]} & \nodata & \nodata \\
HD 201091 & 5.21 & 3.50 & 4441 & \nodata & \nodata & \nodata & 1 & 1 & \nodata & (1) & 1, \textit{[1]} & \nodata & 4, \textit{[1]} \\
HD 201092 & 6.03 & 3.50 & 4107 & \nodata & \nodata & \nodata & 1 & \nodata & \nodata & \nodata & 1, \textit{[1]} & \nodata & 1 \\
HD 202560 & 6.68 & 3.97 & 3599 & \nodata & \nodata & \nodata & 1 & \nodata & 2 & \nodata & (1), \textit{[2]} & 2 & 1 \\
HD 203608 & 4.22 & 9.26 & 6095 & 1 & \nodata & 1 & 1 & \nodata & \nodata & \nodata & 1, \textit{[2]} & \nodata & \nodata \\
HD 20630 & 4.85 & 9.27 & 5709 & \nodata & 1 & 1 & \nodata & 2 & \nodata & 1 & 3 & \nodata & 5 \\
HD 20766 & 5.54 & 12.04 & 5710 & 1 & \nodata & 1 & 1 & \nodata & \nodata & \nodata & 1, \textit{[1]} & \nodata & \nodata \\
HD 20794 & 4.27 & 6.04 & 5432 & 1 & \textit{[1]} & \textit{[1]} & 1 & \nodata & (1) & \nodata & 1, (2) & 1, (1) & 3 \\
HD 20807 & 5.23 & 12.04 & 5847 & \nodata & \nodata & \nodata & 1 & \nodata & \nodata & \nodata & 1, (2) & \nodata & \nodata \\
HD 209100 & 4.69 & 3.64 & 4641 & 1 & \nodata & \nodata & \nodata & \nodata & \nodata & \nodata & 1, (1), \textit{[1]} & 1 & 2, \textit{[1]} \\
HD 210302 & 4.92 & 18.45 & 6364 & \nodata & \nodata & \nodata & (1) & \nodata & \nodata & \nodata & \textit{[1]} & \nodata & \nodata \\
HD 212330 A & 5.31 & 20.32 & 5660 & \nodata & \nodata & \nodata & (1) & \nodata & \nodata & \nodata & \textit{[1]} & \nodata & \nodata \\
HD 216803 & 6.48 & 7.60 & 4601 & \nodata & \nodata & \nodata & \nodata & \nodata & \nodata & \nodata & (1), \textit{[1]} & 1 & \nodata \\
HD 217987 & 7.39 & 3.29 & 3676 & \nodata & \nodata & \nodata & 1 & \nodata & 2 & \nodata & \textit{[2]} & \nodata & 3 \\
HD 219134 & 5.57 & 6.54 & 4874 & \nodata & \nodata & 1 & 1 & \nodata & \nodata & \nodata & (1), \textit{[1]} & \nodata & 3 \\
HD 22484 & 4.30 & 13.92 & 5996 & 1 & \nodata & \nodata & 1 & 1 & (1) & \nodata & (1), \textit{[1]} & \nodata & 1 \\
HD 23249 & 3.54 & 9.09 & 5045 & 1 & \nodata & 1 & \nodata & \nodata & \nodata & \nodata & \textit{[2]} & \nodata & 3 \\
HD 26965 A & 4.43 & 5.01 & 5133 & 1 & 1 & \nodata & (1) & 1 & \nodata & \nodata & \nodata & (1) & 2, \textit{[1]} \\
HD 30495 & 5.50 & 13.23 & 5833 & 1 & \nodata & \nodata & 1 & 1 & \nodata & \nodata & (1), \textit{[1]} & 1, (1) & \nodata \\
HD 30652 & 3.19 & 8.02 & 6443 & 1 & \nodata & \nodata & 1 & 1 & \nodata & \nodata & (1), \textit{[3]} & \nodata & 2 \\
HD 32147 & 6.21 & 8.84 & 4810 & 1 & \nodata & 1 & 1 & \nodata & \nodata & \nodata & (2) & \nodata & 3 \\
HD 33262 A & 4.71 & 11.69 & 6158 & 1 & \nodata & \nodata & (1) & 1 & \nodata & \nodata & 2 & \nodata & 3 \\
HD 34411 & 4.71 & 12.56 & 5854 & \nodata & \nodata & \nodata & 1 & \nodata & \nodata & \nodata & 1, \textit{[2]} & \nodata & 1 \\
HD 35296 & 5.00 & 14.57 & 6131 & 1 & \nodata & \nodata & \nodata & 1 & \nodata & \nodata & 1, (1) & \nodata & 3 \\
HD 3651 A & 5.88 & 11.10 & 5203 & \nodata & \nodata & \nodata & (1) & \nodata & \nodata & \nodata & (2) & 1, (1) & 2 \\
HD 37394 & 6.23 & 12.26 & 5226 & \nodata & \nodata & \nodata & 1 & \nodata & \nodata & \nodata & (1), \textit{[2]} & 1 & 3 \\
HD 38392 & 6.15 & 8.89 & 5027 & 1 & \nodata & \nodata & 1 & \nodata & (1) & \nodata & (1), \textit{[2]} & \nodata & \nodata \\
HD 38393 & 3.60 & 8.90 & 6313 & 1 & \nodata & \nodata & 1 & \nodata & (1) & \nodata & \textit{[2]} & \nodata & \nodata \\
HD 39091 & 5.67 & 18.29 & 5982 & \nodata & 1 & \textit{[1]} & 1 & \nodata & \nodata & \nodata & \nodata & 1 & 4 \\
HD 43042 & 5.19 & 21.78 & 6533 & 1 & \nodata & \nodata & 1, (1) & \nodata & (1) & \nodata & \textit{[1]} & \nodata & 1 \\
HD 43386 & 5.04 & 19.58 & 6525 & 1 & \nodata & \nodata & 1 & \nodata & \nodata & \nodata & \nodata & \nodata & \nodata \\
HD 4391 & 5.79 & 15.04 & 5887 & 1 & \nodata & \nodata & \nodata & \nodata & \nodata & \nodata & \nodata & 1 & \nodata \\
HD 4614 A & 3.44 & 5.92 & 5907 & \nodata & \nodata & \nodata & (1) & \nodata & \nodata & \nodata & (1), \textit{[2]} & \nodata & 1 \\
HD 4628 & 5.74 & 7.43 & 5007 & \nodata & \nodata & \nodata & 1 & 1 & \nodata & \nodata & (1) & 1 & \nodata \\
HD 4813 & 5.19 & 15.92 & 6208 & \nodata & \nodata & \nodata & 1 & \nodata & (1) & \nodata & \textit{[1]} & \nodata & 2 \\
HD 53705 & 5.28 & 17.06 & 5790 & \nodata & \nodata & \nodata & \nodata & \nodata & \nodata & \nodata & \textit{[1]} & \nodata & \nodata \\
HD 55575 & 5.56 & 16.84 & 5902 & \nodata & 1 & \nodata & \nodata & \nodata & \nodata & \nodata & \textit{[1]} & 1 & \nodata \\
HD 693 & 4.89 & 18.88 & 6190 & \nodata & \nodata & \nodata & (1) & \nodata & \nodata & \nodata & \textit{[2]} & \nodata & \nodata \\
HD 69897 & 5.13 & 18.21 & 6269 & \nodata & \nodata & \nodata & \nodata & \nodata & \nodata & \nodata & \textit{[2]} & \nodata & \nodata \\
HD 72905 & 5.64 & 14.43 & 5893 & \nodata & \nodata & 1 & \nodata & 1 & (1) & 1 & 1, (2) & 1 & 2 \\
HD 7570 & 4.96 & 15.25 & 6110 & 1 & \nodata & \textit{[1]} & 1 & \nodata & \nodata & \nodata & \textit{[2]} & 1, (1) & \nodata \\
HD 82885 A & 5.34 & 11.23 & 5518 & 1 & \nodata & 1 & (1) & 1 & \nodata & \nodata & \textit{[3]} & 1 & \nodata \\
HD 84117 & 4.94 & 14.95 & 6163 & \nodata & \nodata & \nodata & 1 & \nodata & \nodata & \nodata & 1, \textit{[2]} & \nodata & \nodata \\
HD 84737 & 5.10 & 18.80 & 5893 & \nodata & \nodata & \nodata & \nodata & \nodata & \nodata & \nodata & (1), \textit{[1]} & 1 & \nodata \\
HD 86728 A & 5.38 & 14.92 & 5743 & \nodata & \nodata & \nodata & \nodata & \nodata & \nodata & \nodata & 1 & 1 & \nodata \\
HD 88230 & 6.61 & 4.87 & 4097 & \nodata & \nodata & \nodata & \nodata & \nodata & \nodata & \nodata & (1), \textit{[2]} & 2 & \nodata \\
HD 90839 & 4.82 & 12.94 & 6164 & \nodata & \nodata & \nodata & 1 & \nodata & \nodata & \nodata & (2), \textit{[1]} & 1 & \nodata \\
HD 95128 & 5.03 & 13.88 & 5880 & \nodata & \nodata & 1 & 1 & \nodata & 2 & \nodata & 1, \textit{[2]} & 1 & 2 \\
HD 95735 & 7.52 & 2.55 & 3469 & 1 & \nodata & 1 & \nodata & \nodata & 2 & \nodata & \nodata & 2 & 1, \textit{[2]} \\
\end{longtable}

\end{ThreePartTable}

\subsection{Spectral Coverage Overview}

\begin{figure}
    \centering
    \includegraphics[width=0.4\linewidth]{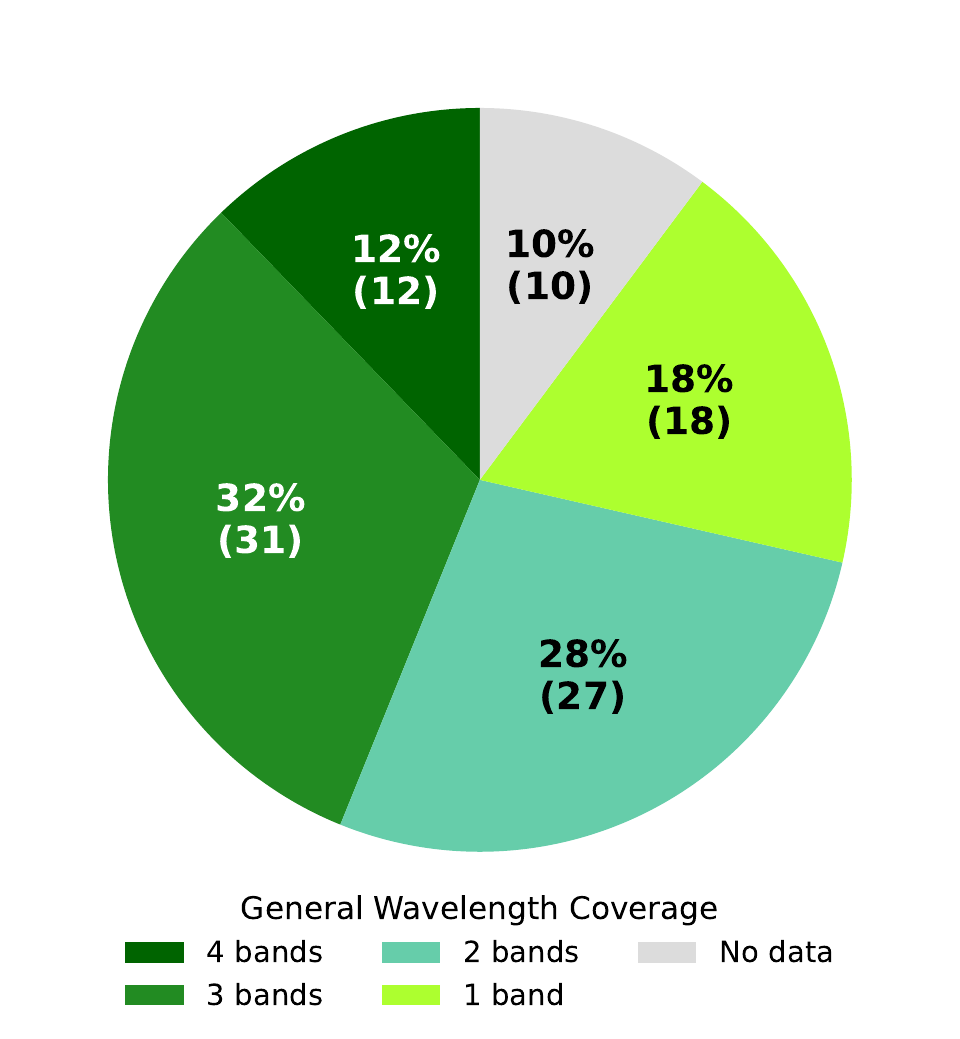}
    \includegraphics[width=0.4\linewidth]{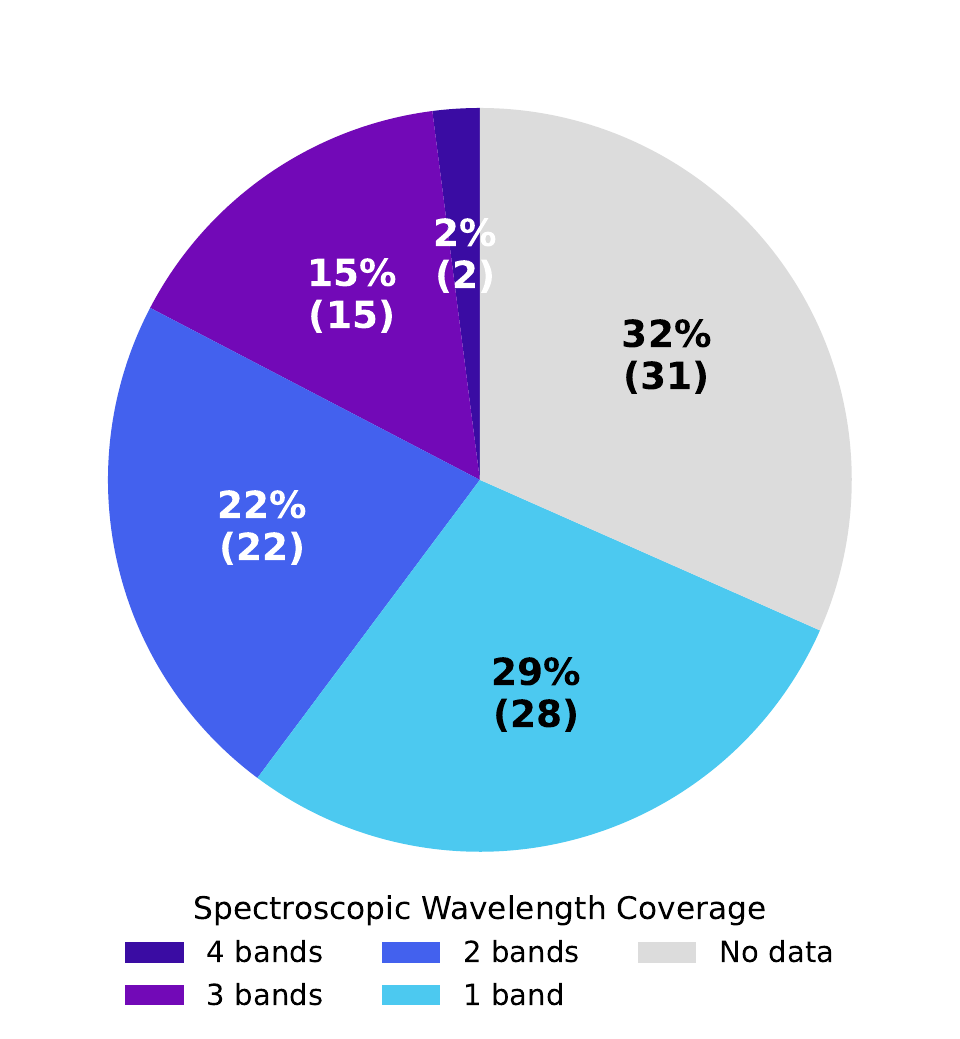}
    \caption{\textit{Left:} Distribution of stars by the number of wavelength bands (X-ray, EUV, FUV, NUV) with usable observations (either spectroscopy or photometry). Each star is counted once based on how many distinct bands it has good data in. The ``No data” category captures stars that lack usable observations in any of the four bands, this includes stars that are limited to data with ``known issues". Both the percentage of the full target sample and the total number of stars are indicated within each wedge. \textit{Right:} The distribution of stars limited to spectroscopy.}
    \label{fig:pie_bands}
\end{figure}

The overall distribution of usable-quality data across different portions of the high energy spectrum is summarized in Figure \ref{fig:pie_bands}. These two pie charts categorize target stars by the number of distinct wavelength bands (X-ray, EUV, FUV, NUV) in which data of satisfactory quality are available, either with both photometric and spectroscopic observations or limiting to just spectroscopy. Only 12\% (12 stars) of the sample is well characterized, having either spectroscopic or photometric observations in all four bands: HD~114710, HD~131156 A, HD~147513, HD~166, HD~17925, HD~201091, HD~20630, HD~26965 A, HD~30652, HD~33262 A, HD~35296, and HD~72905. Just 2\% (2 stars) have usable-quality spectroscopic observations in all four bands: HD~131156 A ($\xi$~Boo A) and HD~20630 ($\kappa$ Cet). This is driven by the dearth of EUVE spectra.  If we only consider X-ray, FUV, and NUV spectroscopy, there are an additional 15\% (or 15 stars) with good data coverage. Twenty-nine percent (28 stars) of the stellar sample only has NUV spectroscopy available, while another 32\% (31 stars) lacks any reliable high energy spectroscopic data. Ten percent of the sample (10 stars) lacks usable data in any band: HD~134083, HD~156274 A, HD~210302, HD~212330 A, HD~46588 A, HD~53705, HD~58855, HD~65907, HD~693, and HD~69897. 

\begin{figure}
    \centering
    \includegraphics[width=0.9\linewidth]{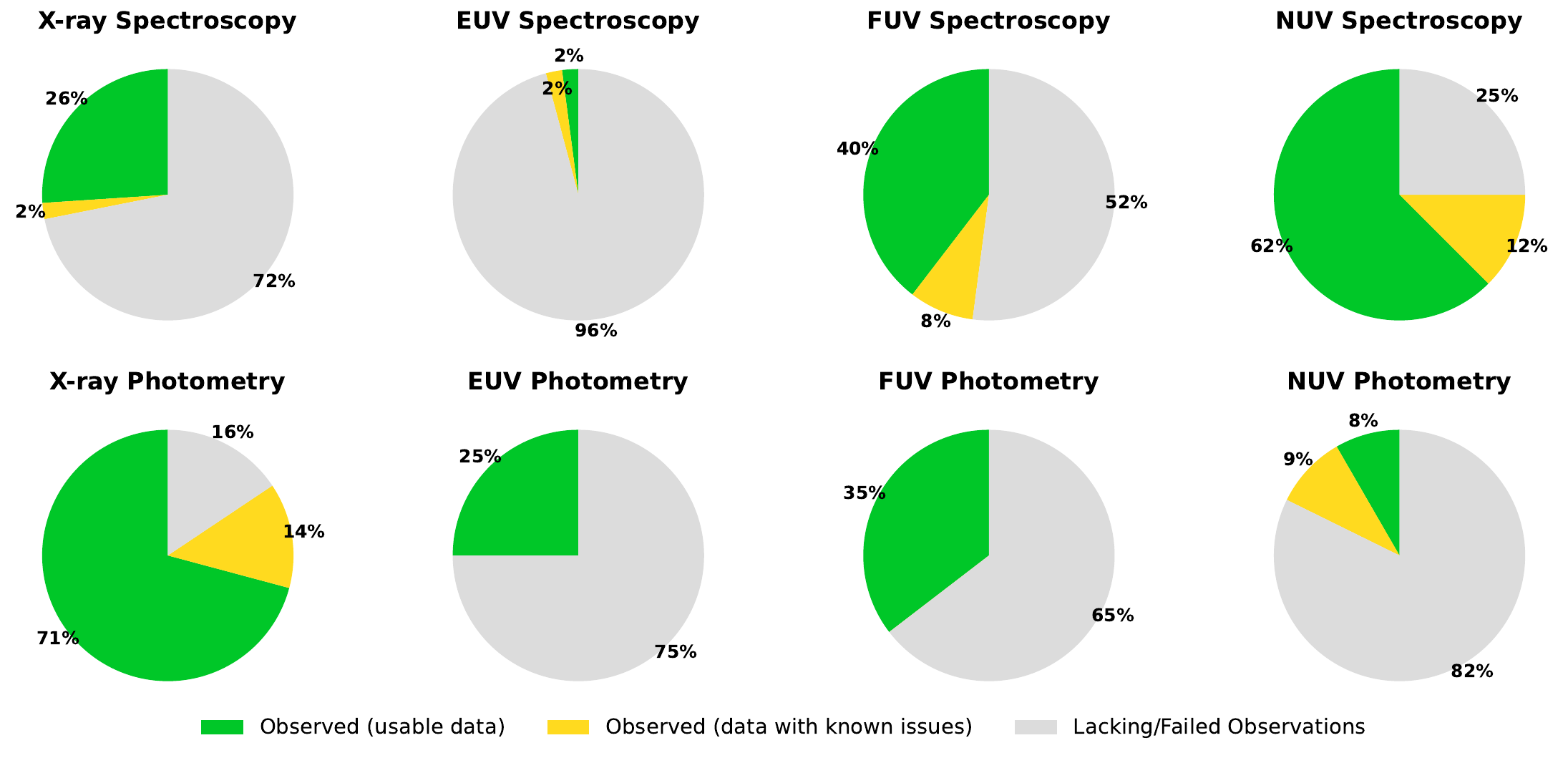}
    \caption{Data availability distributions across four wavelength bands (X-ray, EUV, FUV, NUV) and two data types (spectroscopy and photometry). Each pie chart shows the percentage of the full 98-star sample that falls into three categories: usable-quality data (green), data with known issues (yellow), and missing or failed observations (gray). The best available quality per star is used for each band+type combination.}
    \label{fig:pie_type}
\end{figure}

Further clarification is provided by Fig.~\ref{fig:pie_type}, which dissects the dataset by displaying the availability of usable-quality observations within each wavelength band and for each data type (spectroscopy vs. photometry). This visualization elucidates differences in spectral resolution and instrumental capabilities among various bands. The distribution of usable-quality and lower-quality observations varies significantly across wavelength bands and data types. The most complete coverage comes from X-ray photometry and NUV spectroscopy, with 71\% and 62\% of targets having usable-quality data, respectively. These two categories dominate the high-energy dataset and are likely to anchor any broad statistical inferences about the stellar high-energy environment. In contrast, EUV spectroscopy and NUV photometry stand out for their poor coverage. As previously mentioned, only 2\% of stars have usable EUV spectra, and just 8\% have usable NUV photometry; both fall below the 10\% threshold, making them unsuitable for sample-wide trends. Other bands, such as FUV spectra (40\%) and FUV photometry (35\%), contribute moderate coverage, but still fall short of being representative for the full sample. Notably, X-ray spectroscopy data is more limited, with only 26\% of targets having usable-quality spectra, and 2\% flagged as likely problematic.

\subsection{Observatory-Specific Data Availability and Sample Demographics}

\begin{figure}
    \centering
    \includegraphics[width=0.9\linewidth]{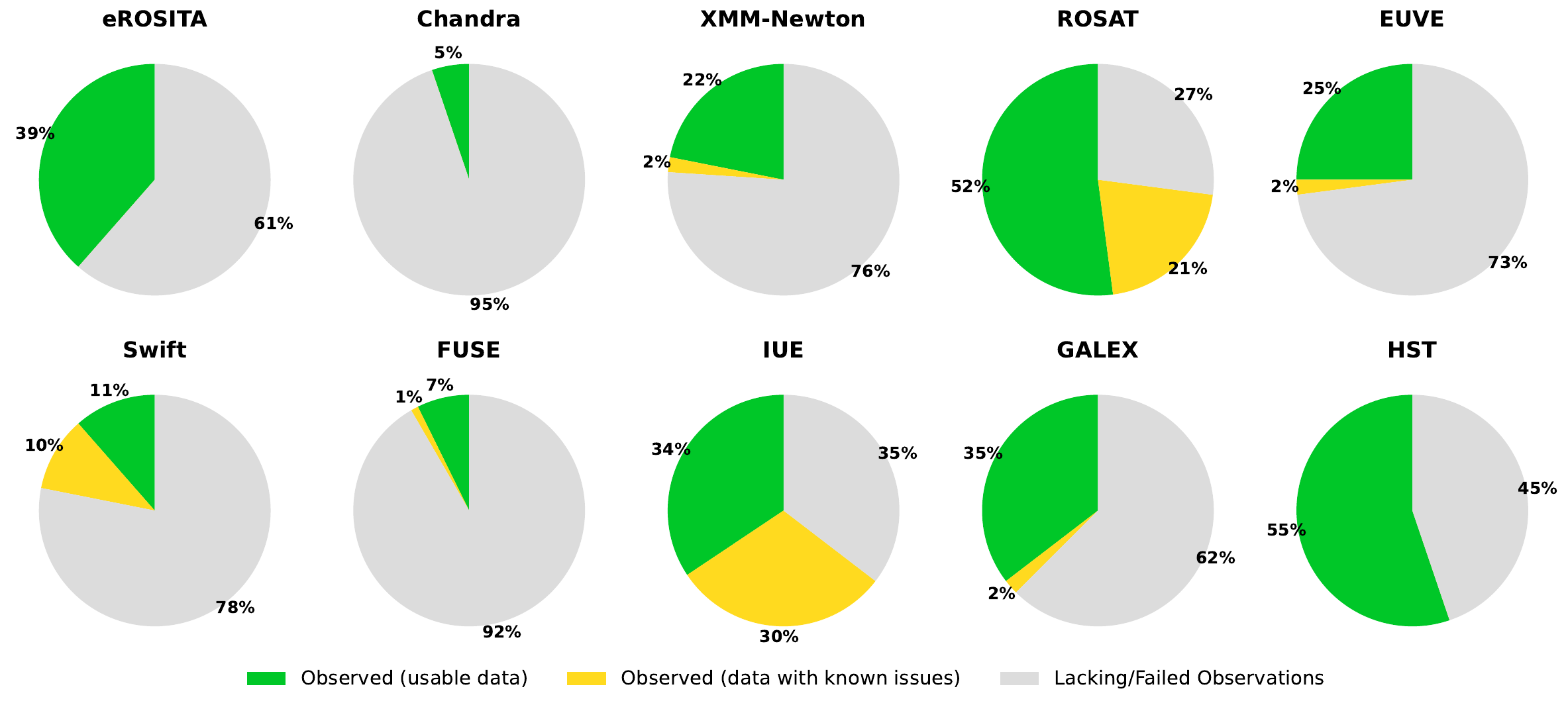}
    \caption{Breakdown of data availability for the target stars by observatory. Each pie chart shows the fraction of stars with usable-quality data (green), data with known issues (yellow), and missing or failed observations (gray), normalized over the full sample of 98 stars. For each observatory, the best available data quality per star is used when multiple observations exist.}
    \label{fig:pie_observatory}
\end{figure}

The contributions of individual observatories are examined in Figure \ref{fig:pie_observatory}, which features a set of pie charts detailing the fraction of target stars observed by each facility. This figure highlights the distribution of observational coverage and data quality contributed by different facilities. For example, looking at the combined number of stars with usable-quality data (green) and data with known issues (yellow), it is clear that ROSAT and IUE observed the highest total percentages of targets; however, these observations are also the most prone to issues and our initial vetting of the data quality suggest that at least 20--30\% of the data is unreliable. Slightly more than half of the stars (55\%) have either HST FUV and/or NUV spectra available, while just 31\% (31 stars) have HST spectra in both. The other seven observatories have observed less than 40\% of the targets, with FUSE, Chandra, and Swift yielding the least amount of usable-quality data for this stellar sample.

\begin{figure}
    \centering
    \includegraphics[width=1.0\linewidth]{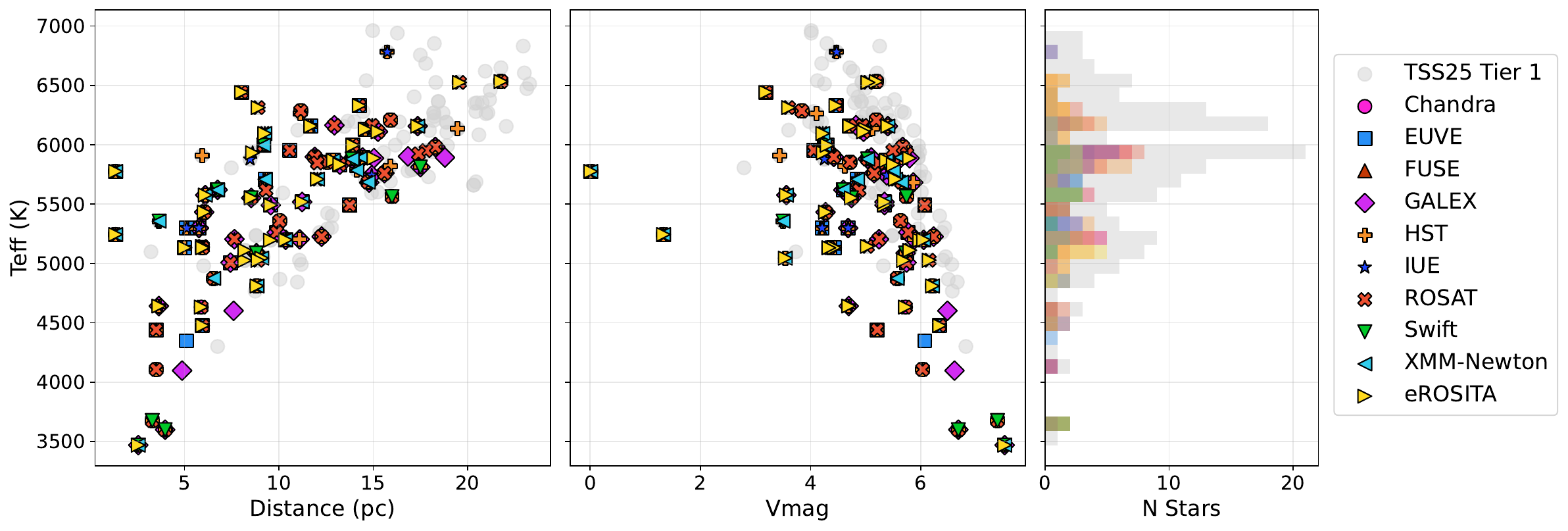}
    \caption{Demographics of stars with usable-quality observations from each observatory, compared to the full TSS25 Tier 1 target list. All panels show stellar effective temperature ($T_{\mathrm{eff}}$) on the $y$-axis. \textit{Left:} $T_{\mathrm{eff}}$ vs. distance. \textit{Middle:} $T_{\mathrm{eff}}$ vs. $V$-band magnitude. \textit{Right:} Histogram of the number of stars per $T_{\mathrm{eff}}$ bin. Different observatories are indicated by distinct colors and symbols, and stack vertically if a star has usable-quality data from multiple observatories. Histogram bars start at zero and show the total number of stars in each $T_{\mathrm{eff}}$ bin. Versions of this figure with each observatory shown individually is provided in the Appendix.}
    \label{fig:full_demo}
\end{figure}

Figure \ref{fig:full_demo} presents demographic comparisons between the full TSS25 Tier 1 target list (164 stars) and the subset of stars considered in our analysis (98 stars) with usable-quality data from each observatory. We also present versions with each observatory shown individually in the Appendix. These comparisons help reveal potential selection effects related to stellar type, brightness, and proximity, as well as any biases associated with the observational capabilities of individual observatories. Looking at the y-axis of Figure \ref{fig:full_demo}, we find comprehensive coverage across the full temperature range of the stellar sample. While the absolute number of hotter stars targeted in our subsample, (ExEP list Tiers A and B) is comparable to that of cooler stars, their relative fraction is smaller due to the larger number of hot stars in the overall TSS25 Tier 1 sample (ExEP list Tiers A, B, and C). Tier C contributes substantially to this as it skews hotter as a full population; for example, Tier C is comprised of 37 F-type stars compared to 14 and 15 in Tiers A and B, respectively. 

The left panel of Figure \ref{fig:full_demo} highlights how our 98-star sub-sample is limited to within 21.7\,pc, while the full TSS25 Tier 1 list includes more distant targets. The sub-sample is densest between $\sim$5--20\,pc, with a slight tendency for the closest stars to be cooler, consistent with the nearby abundance of M dwarfs. Observatories such as ROSAT and eROSITA contribute coverage across the full distance range, while others, especially EUVE and FUSE, were more tightly distance-limited, typically to within 15\,pc, as shown in Appendix Figures \ref{fig:a1}-\ref{fig:a4}.

The $V$-band magnitude distribution (middle panel) further illustrates how observational constraints favor optically bright targets. Most stars with usable data fall between $V$ = 4\,mag and 7\,mag, with only a few very bright ($V <$ 2\,mag) or faint ($V >$ 7\,mag) stars included. Observatory-specific limits are apparent here as well: for example, Swift, EUVE, and GALEX preferentially observed brighter stars, while ROSAT and eROSITA include fainter ones. Outliers at either extreme are generally associated with a single observatory, often due to specific targeting decisions or survey sensitivities.

The histogram in the right panel reflects these selection effects, showing a strong peak in coverage for G-type stars (5000--6000\,K) and three bright M dwarfs. A secondary group of hotter stars ($\sim$6000--6300\,K) is also present, primarily due to the broader reach of ROSAT and eROSITA. However, several observatories (including Chandra, FUSE, GALEX, Swift, and XMM-Newton) are notably underrepresented or missing entirely above $\sim$6000\,K, pointing to temperature-dependent biases, likely tied to specific science goals or instrumental limitations.

Together, these trends demonstrate how stellar demographics in the archival UV and X-ray sample are shaped by a combination of mission capabilities, observing strategy, and proximity-driven brightness biases. These effects must be considered in any statistical or comparative analysis of the sample.

\section{Estimating NUV Fluxes and ISM Attenuation for TSS25}

A comprehensive understanding of the high-energy radiation environments of exoplanet host stars is critical for designing instruments like those planned for HWO. Ultraviolet and X-ray observations provide key insights into atmospheric heating, mass loss, and photochemical evolution, but as shown in Sect.~\ref{sec:obs_coverage}, existing data are incomplete and biased. Most archival UV and X-ray observations are shaped by instrumental sensitivity limits, targeted observing strategies, and attenuation from the ISM. These limitations hinder efforts to uniformly characterize the full population of potential HWO targets.

In our curated 98-star subsample of the TSS25 list, only 62\% of targets have usable NUV observations, primarily limited to the brightest and nearest stars. Even within Tier 1, just 37\% of stars (36 total) have sufficiently high-resolution NUV (e.g., STIS E230H of Mg~{\sc ii}) or FUV (e.g., STIS E140H or E140M of \lya) spectra to enable precise determination of interstellar hydrogen column densities. Because of the complex and structured nature of the local ISM within 15 pc \citep{Redfield2008, Youngblood2025}, direct line-of-sight measurements are strongly preferred. Yet for most targets, such data are unavailable, requiring model-based or interpolated estimates from large-scale ISM maps.

To overcome these observational gaps and enable consistent treatment of the entire TSS25 sample, we extend our analysis beyond archival data. We use synthetic stellar atmosphere models to estimate narrowband NUV fluxes for all $\sim$13,000 TSS25 stars and derive $N$(H~{\sc i}) column densities for every Tier 1 target. As discussed in Sect.~1, these modeled quantities are vital for forward modeling of instrument performance, including coronagraph exposure time calculations and correction of UV line fluxes for ISM attenuation \citep[e.g.,][]{Wood2005, Frisch2011, Youngblood2025}. The following sections describe our methodology for generating these estimates and examine their implications for HWO planning across the target sample.

\subsection{PHOENIX Predictions of NUV Fluxes}


Representative narrow-band NUV photometric estimates (2500--4500\,\AA) were generated for the full TSS25 list\footnote{Combined, Tiers 1, 2, and 3 of the TSS25 target list consists of all $\sim$13,000 nearby, bright stars in the HPIC which would be potential HWO targets. This input catalog is agnostic about assumptions of HWO's mission design, and is a required input for trade studies and yield calculations. However, most of the stars in the HPIC will likely not be good targets for a specified HWO design, and less precise precursor knowledge of stellar properties is required for this broad population.} at the request of the \textit{Exoplanet Science Yields} group (\textit{private communication}) to support exoplanet yield and coronagraph exposure time analyses. These estimates were derived using BT-Settl PHOENIX stellar atmosphere models \citep{Allard2012}, which extend into the UV but do not include prescriptions for chromospheric or transition region emission. An additional modeling limitation is the use of local thermodynamic equilibrium (LTE), which can lead to overestimation of emission line strengths and thus artificially high NUV continuum levels, whereas UV emission lines are more accurately treated under non-LTE conditions.

\begin{figure}
    \centering
    \includegraphics[width=0.6\linewidth]{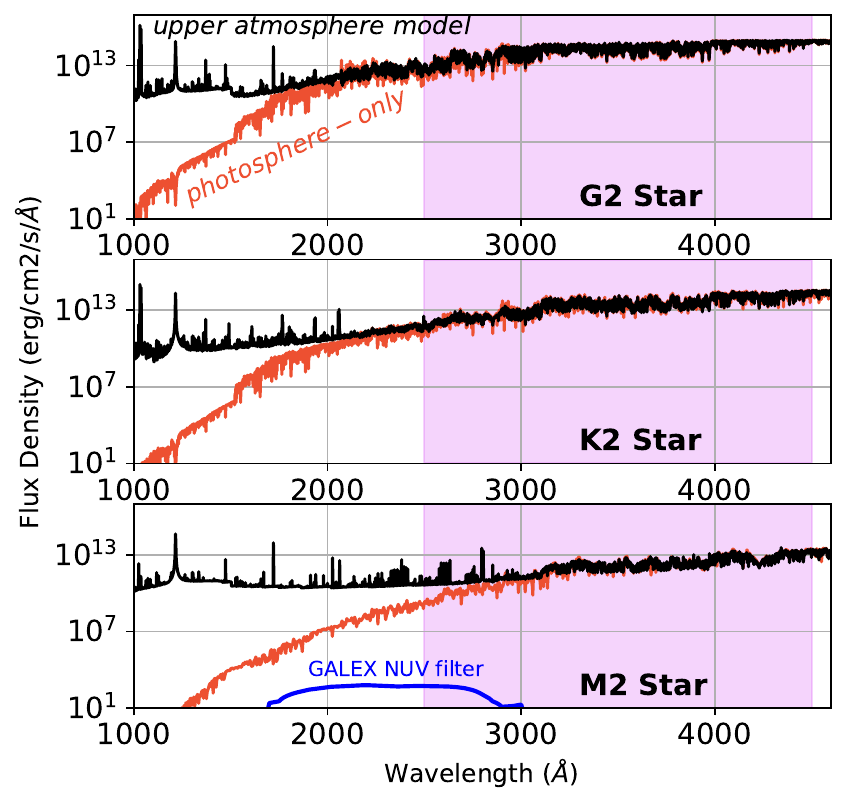}
    \caption{PHOENIX model spectra for G, K, and M stars with (black) and without (red) extended upper atmospheres (i.e., chromosphere and transition region). NUV wavelengths used in HWO yield analyses are highlighted in purple. We plot the GALEX NUV transmission curve in the bottom panel for reference. Models similar to the red spectra were used to estimate NUV fluxes in narrow bands for HPIC catalog stars. These models tend to underpredict UV fluxes, though the effect is more pronounced for M stars (which represent a small fraction of likely HWO targets) and at FUV wavelengths (not relevant for the yield analyses).}
    \label{fig:phx_photo}
\end{figure}

As illustrated in Figure~\ref{fig:phx_photo}, the omission of upper atmospheric layers has important consequences: the synthetic NUV fluxes systematically underestimate the true stellar output in this band, particularly at shorter wavelengths and for later-type stars. While the discrepancy is most dramatic for M dwarfs, the impact is more moderate, and generally within a factor of two, for FGK stars. Given that the vast majority of TSS25 Tier 1 stars are FGK-type (with only three M stars), this limitation does not significantly affect our overall results. In fact, the modeled NUV fluxes can be considered reliable lower limits for the primary mission-relevant targets.

To assign models to each star, we selected the closest-matching BT-Settl PHOENIX spectrum based on a minimum Euclidean distance in $T_{\rm eff}$, surface gravity ($\log{g}$), and [Fe/H], using stellar parameters from the HPIC catalog \citep{Tuchow2024}. Model NUV fluxes were then computed via flat integration over 500\,\AA-wide bins spanning 2500--4500\,\AA. Because all model fluxes are defined at the stellar surface, they were scaled to Earth by a factor of $(R\star^2 / d^2)$, where $R_\star$ is the stellar radius and $d$ is the distance \citep[Zenodo repository available at ][]{PeacockZenodo2025}.

To evaluate the accuracy of the synthetic model fluxes, we compared them to observed GALEX and HST NUV photometry. Synthetic GALEX-band fluxes were calculated by convolving both the model and observed HST spectra with the GALEX NUV filter response (1750--2800\,\AA) to ensure consistency. HST spectra from the LOWLIB catalog, which spans the entire GALEX NUV bandpass (1710--10070\,\AA), served as a reliable observational baseline. However, only three targets in our sample have robust, unsaturated GALEX NUV measurements. To increase sample size, we also included non-linear regime GALEX observations that were empirically corrected using white dwarf calibration curves \citep{Wall2019}, though these corrections introduce considerable uncertainty and are classified as ``data with known issues'' in the rest of this paper.

\begin{figure}
    \centering
    \includegraphics[width=0.85\linewidth]{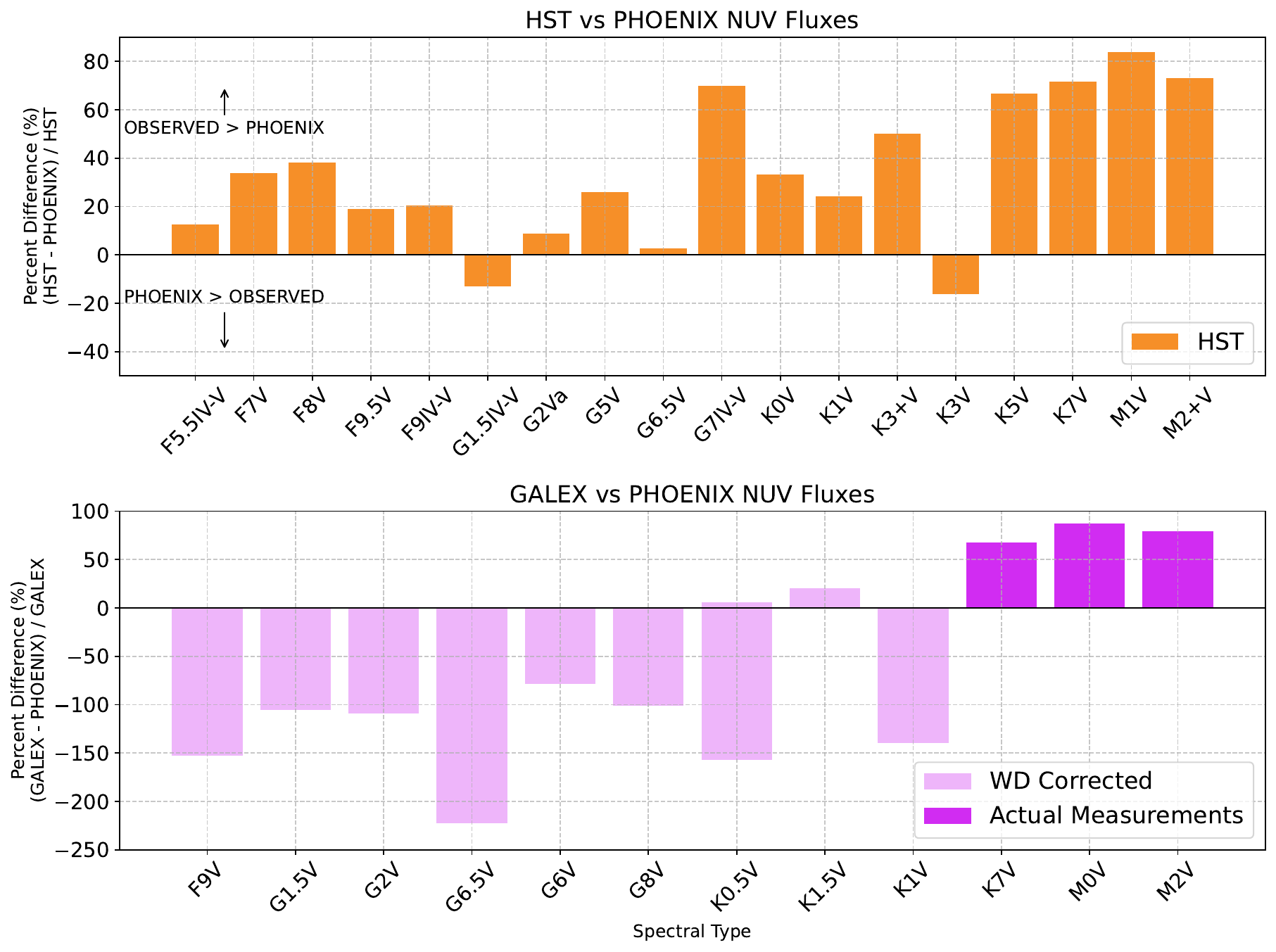}
    \caption{\textit{Top:} When compared to HST NUV flux measurements, the BT-Settl PHOENIX models nearly always underpredict NUV fluxes for FGKM stars, with the discrepancy increasing with decreasing temperature. In all cases the model predicted fluxes are within a factor of two. The two instances where the model flux exceeds the observations coincide with HST spectra that have excessive noise and gaps shortward of 2100\,\AA. \textit{Bottom:} When compared to GALEX flux measurements, the BT-Settl PHOENIX models underpredict NUV flux for M stars at levels similar to the HST data, but predict larger flux values for FGK stars. All but one of the FGK stars in this plot fall within the non-linear regime in the NUV GALEX detectors, requiring a white dwarf correction, which may yield significant uncertainty. 
}
    \label{fig:phx_v_obs}
\end{figure}

In Figure \ref{fig:phx_v_obs}, the comparisons reveal that the BT-Settl PHOENIX models underpredict the GALEX NUV-band flux by approximately 40\% for F and G stars, 50\% for K stars, and up to 70\% for M dwarfs relative to HST-based measurements. Comparisons to GALEX observations show a similar underprediction for M stars, but apparent overpredictions for FGK stars, likely due to the saturation and correction uncertainties in GALEX data. These discrepancies underscore the limited utility of GALEX for validating modeled NUV fluxes, particularly for the bright nearby stars most relevant to HWO.

Taking all comparisons into account, we estimate that for the vast majority of stars in TSS25, the model-predicted NUV fluxes represent conservative lower limits, typically within a factor of two of the intrinsic values. These fluxes are therefore well-suited for prioritizing targets and estimating minimum exposure times for HWO. All modeled NUV fluxes have been incorporated into the HPIC catalog\footnote{The HPIC catalog is available at \url{https://emac.gsfc.nasa.gov/?cid=2403-004}} and the modeled values for all TSS25 Tier 1 stars are also provided in Table \ref{tab:nuv_fluxes} for reference.

\begin{deluxetable*}{lrrrrrrrrrrrr}
\tablecaption{Modeled NUV Fluxes for TSS25 Tier 1 Stars\label{tab:nuv_fluxes}}
\tablehead{
\colhead{Name} &
\colhead{$T_{\mathrm{eff}}$} &
\colhead{$\log g$} &
\colhead{[Fe/H]} &
\colhead{$d$} &
\colhead{Radius} &
\colhead{$F_{\mathrm{G,NUV}}$} &
\colhead{$F_{2500-4000}$} &
\colhead{$F_{2000-2500}$} &
\colhead{$F_{2500-3000}$} &
\colhead{$F_{3000-3500}$} &
\colhead{$F_{3500-4000}$} &
\colhead{$F_{4000-4500}$} \\
\colhead{} &
\colhead{(K)} &
\colhead{(dex)} &
\colhead{(dex)} &
\colhead{(pc)} &
\colhead{($R_\odot$)} &
\multicolumn{7}{c}{log$_{10}$(Flux) [erg s$^{-1}$ cm$^{-2}$]}
}
\startdata
	HD 100623 A & 5196 & 4.62 & -0.40 & 9.56 & 0.8 & -12.86 & -11.47 & -13.13 & -12.19 & -11.42 & -11.22 & -10.93 \\
	HD 101501 & 5491 & 4.50 & -0.05 & 9.57 & 0.9 & -12.33 & -11.13 & -12.64 & -11.64 & -11.06 & -10.94 & -10.65 \\
	HD 102365 & 5618 & 4.41 & -0.31 & 9.32 & 1.0 & -11.92 & -10.86 & -11.98 & -11.36 & -10.76 & -10.71 & -10.41 \\
	HD 10360 & 5025 & 4.59 & -0.24 & 8.19 & 0.8 & -13.08 & -11.58 & -13.58 & -12.30 & -11.50 & -11.37 & -10.95 \\
	HD 10361 & 5111 & 4.63 & -0.19 & 8.20 & 0.7 & -13.00 & -11.56 & -13.46 & -12.24 & -11.48 & -11.35 & -10.97 \\
	HD 10476 & 5204 & 4.51 & -0.04 & 7.64 & 0.8 & -12.62 & -11.24 & -13.04 & -11.87 & -11.17 & -11.04 & -10.69 \\
	HD 10700 & 5356 & 4.56 & -0.51 & 3.65 & 0.8 & -11.59 & -10.37 & -11.81 & -11.02 & -10.28 & -10.18 & -9.88 \\
	HD 10780 & 5358 & 4.50 & 0.03 & 10.04 & 0.8 & -12.54 & -11.28 & -12.88 & -11.83 & -11.21 & -11.10 & -10.78 \\
	HD 109358 & 5878 & 4.37 & -0.20 & 8.47 & 1.1 & -11.53 & -10.54 & -11.69 & -10.95 & -10.48 & -10.38 & -10.16 \\
	HD 114710 & 5996 & 4.36 & 0.06 & 9.20 & 1.1 & -11.44 & -10.51 & -11.56 & -10.91 & -10.45 & -10.35 & -10.15 \\
\vdots & \vdots & \vdots & \vdots &\vdots &\vdots & \vdots & \vdots & \vdots &\vdots &\vdots & \vdots & \vdots \\
\enddata
\tablecomments{
Estimated NUV fluxes for TSS25 Tier 1 stars computed using BT-Settl PHOENIX stellar atmosphere models. Stellar parameters are adopted directly from the HPIC catalog. All \textbf{fluxes are taken from the stellar surface and scaled by $(R_\star/d)^2$ to represent the flux that would be measured at Earth}. The GALEX NUV fluxes ($F_{\mathrm{G,NUV}}$) are calculated by convolving the model spectra with the GALEX NUV transmission curve (shown in Figure~\ref{fig:phx_photo}), while fluxes in the other wavelength bands (listed in $\AA$) are computed from flat (boxcar) integrations. This table is published in its entirety in machine-readable format in the online journal.
}
\end{deluxetable*}

\subsection{LISM Estimates}

Accurate reconstruction of the high-energy stellar spectrum incident on exoplanets requires not only detailed stellar modeling but also consideration of absorption by the LISM. This is particularly important in the EUV, where interstellar H~{\sc i} and He~{\sc i} absorption can obscure nearly all of the intrinsic stellar emission, even for nearby stars. As a result, LISM absorption characterization in other wavelength bands, such as the NUV and FUV, is critical for evaluating the intrinsic EUV flux reaching orbiting planets. 

For the full TSS25 Tier 1 target list (164 stars), we estimated H~{\sc i} column densities using the two-dimensional H~{\sc i} map of the LISM presented in \citealt{Youngblood2025}. Estimates were obtained via the publicly available \texttt{LISM\_NHI} tool\footnote{\url{https://github.com/allisony/LISM_NHI}} (version 1.0), which returns $\log N$(H~\textsc{I}) values based on the target's right ascension, declination, and distance. For the target stars, the estimated column densities span $\log N$(H~\textsc{I}) = 17.61--18.80, with an average value of 18.10 (Table \ref{tab:ism})\footnote{Given average correlations between H column densities and reddening among field stars, e.g. $N$(H~{\sc i})/$E(B-V) = 4.93(\pm0.28) \times 10^{21}$ atoms\,cm$^{-2}$\,mag$^{-1}$ \citep{Diplas1994}, the measured column densities suggest that the HWO Tier 1 target stars suffer from negligible reddening due to dust, approximately $E(B-V) \simeq$ 0.083--1.28\,mmag. For typical Galactic ISM dust $A_{V}/E(B-V)$ $\simeq$ 3.1, this is consistent with Johnson $V$-band extinction of $A_{V}$ $\simeq$ 0.00026--0.0040\,mag. Hence reddening and extinction can be safely ignored in visible and near-IR SED modeling of the HWO targets, given typical photometric uncertainties ($>$1\%) and absolute calibration uncertainties ($\sim$1-2\%). However the effects are somewhat larger in the NUV and FUV bandpasses \citep{Wall2019}, $A_{\rm NUV}$ $\simeq$ 0.0006--0.009\,mag and $A_{\rm FUV}$ $\simeq$ 0.0007--0.01\,mag.} We note that many sight lines have multiple absorbers, and these values represent the total integrated column along the line of sight. 

From the 98-star subset used for our archival data analysis, we report measured $\log N$(H~{\sc i}) values for stars with published literature measurements. A total of 34 stars have measured values, derived either directly from H~{\sc ii} observations or scaled from Mg~{\sc ii} or Fe~{\sc ii}, as noted in Table~\ref{tab:ism}. We also flag five stars with available high-resolution spectra that could be used to measure $\log N$(H~{\sc i}), but which have not yet been analyzed.

\begin{deluxetable}{lccccccc}[htb!]
\tablecaption{LISM Column Densities for TSS25 Tier 1 Stars}
\label{tab:ism}
\setlength{\tabcolsep}{5pt}
    \tablehead{
    \colhead{Star Name} & \colhead{RA} & \colhead{Dec.} & \colhead{$d$ (pc)} & \colhead{Estimated} & \colhead{Measured} & \colhead{Source} &\colhead{Ref.} \\
         &  &  &  & \colhead{$\log N$(H~\textsc{i})} & \colhead{$\log N$(H~\textsc{i})} & &
    }
    \startdata
	HD 100623 A	&	173.6228596	&	-32.83133899	&	9.56	&	17.91	$\pm$	0.22	& \nodata & \nodata& \nodata \\
	HD 101501	&	175.2625664	&	34.20163403	&	9.57	&	17.91	$\pm$	0.22	& \nodata & \nodata & \nodata\\
	HD 102365	&	176.6294666	&	-40.50035555	&	9.32	&	17.85	$\pm$	0.22	& \nodata & \nodata & \nodata\\
	HD 10360	&	24.94922141	&	-56.19331945	&	8.19	&	18	$\pm$	0.22	& \nodata & \nodata & \nodata\\
	HD 10361	&	24.9481872	&	-56.19644756	&	8.2	&	18	$\pm$	0.22	& \nodata & \nodata & \nodata\\
	HD 10476	&	25.62401456	&	20.26851674	&	7.64	&	18.03	$\pm$	0.22	& \nodata & \nodata & \nodata\\
	HD 10700	&	26.01704803	&	-15.93748189	&	3.65	&	18.01	$\pm$	0.22	& 18.01 & H~\textsc{i} & (1) \\
	HD 10780	&	26.93680679	&	63.85250203	&	10.04	&	18.16	$\pm$	0.35	& \nodata & \nodata & \nodata\\
	HD 109358	&	188.4356666	&	41.35747824	&	8.47	&	17.85	$\pm$	0.22	& \nodata & \nodata & \nodata\\
	HD 114710	&	197.9683212	&	27.878183	&	9.2	&	17.85	$\pm$	0.22	& 17.94 & Mg~\textsc{ii} & (2) \\ 
    \vdots    & \vdots    & \vdots & \vdots &\vdots
    \enddata
   \tablerefs{(1) \citet{Wood2005}; (2) \citet{Malamut2014}; (3) \citet{Redfield2002}; (4) \citet{Linsky1996}; (5) \citet{Redfield2004}; (6) \citet{Nisak2025}; (7) \citet{Wood2000b}; (8) \citet{Wood2000a}; (9) \citet{Zachary2018}; (10) \citet{Edelman2019}; (11) \citet{Wood1998}; (12) \citet{Youngblood2022}; (13) \citet{Wood1996}; (14) \citet{Wood2021}; (15) \citet{Vannier2025}; (16) \citet{Wood2014}}
    \tablecomments{Estimated values from the LISM$_{\rm N\,HI}$ tool from \citealt{Youngblood2025}. Measured values for Tier A and B stars are derived directly from H~{\sc i} observations, or scaled from Mg~{\sc ii} or Fe~{\sc ii} observations, as specified in the Source column. Stars marked with $^{\dagger}$ have high-resolution spectra available that could yield measured values, but these spectra have not yet been analyzed.\\  (This table is available in its entirety in machine-readable form in the online article).}
\end{deluxetable}

\section{Future Analysis and Observing Campaigns}

\subsection{Future Analysis}

This work establishes the current UV and X-ray coverage for likely HWO targets, laying the foundation for a broad range of additional analyses enabled by the archival high-energy dataset. Below, we outline several future avenues of study that will inform target prioritization and guide atmospheric interpretation for rocky exoplanets observed with HWO.

\textbf{Identify targets with Earth-like XUV and UV environments in their habitable zones:}\\
Using the available high-energy spectra, it is possible to compute the incident fluxes in the habitable zones of each star and compare them to Earth's current and early XUV and UV environments. This will help to flag systems whose radiation environment may support long-term atmospheric retention or habitability, and distinguish those that may be less favorable due to enhanced atmospheric erosion. According to \citet{Binder2024}, 16 of our target stars have solar-like L$_X$/L$_{\rm bol}$ values in their habitable zones, indicating a similar X-ray energy deposition compared to the Earth-Sun system. However, it has not yet been quantified how the full EUV–NUV spectrum compares across this subset. 

\textbf{Flag outlier stars with exceptionally high or low X-ray and UV fluxes:}\\
Identifying stars with unusually high or low activity levels, compared to expectations based on their spectral types and ages, can highlight potentially anomalous systems. These may be of particular interest for follow-up if they show unexpected atmospheric outcomes on their planets-or warrant caution if they represent poor analogs to typical systems.

\textbf{Compare X-ray and UV fluxes to age and traditional activity indicators:}\\
Many of the HWO target stars have existing measurements of chromospheric activity (e.g., Ca~{\sc ii} H\&K), rotational periods, or estimated ages. Cross-comparing these proxies to direct high-energy measurements will help calibrate age-activity relationships across a range of stellar types, and identify inconsistencies or new trends in low-mass stars.

\textbf{Quantify the time-integrated high-energy exposure by spectral type:}\\
While current observations provide snapshots of stellar activity, understanding the cumulative impact on planetary atmospheres requires integrating high-energy fluxes over stellar lifetimes. This analysis can build on existing empirical or model-based activity decay curves to estimate lifetime XUV doses, which are key inputs to atmospheric evolution models.

\textbf{Assess variability and flare statistics in repeat observations:}\\
Several stars in the sample have been observed multiple times with the same instrument and mode, enabling time-domain studies of high-energy variability. Identifying flare frequency, amplitude, quiescent variability, and more certain relationships between the different wavebands during different types of flares will help contextualize snapshot spectra and improve estimates of average and extreme high-energy conditions.

\subsubsection{Predicting EUV Fluxes from Archival Data}

While both X-ray and EUV flux contribute to atmospheric mass loss, EUV photons are the primary drivers for two key reasons. First, EUV photons are absorbed in the uppermost layers of planetary atmospheres, where radiative cooling is inefficient and heating efficiency is maximized. Second, despite having lower luminosity than X-rays at young ages, the total number of EUV photons emitted by cool stars exceeds the X-ray photon output by a factor of 3–90 across all stellar types \citep{woods2009, fontenla2016,Garcia-Sage2017, king2021}

While stellar X-rays are often readily observable with a variety of instruments like Chandra, XMM-Newton, and eROSITA, unfortunately, direct EUV observations of exoplanet host stars are exceedingly rare. The only EUV-dedicated mission to date, EUVE, acquired spectra of just over a dozen cool main sequence stars, including two of our targets. Despite its limitations, EUVE provided critical insights into the EUV output of nearby stars. Given the importance of the EUV band, particularly the 100--500\,\AA\ region, for modeling photochemistry and escape in planetary atmospheres, renewed investment in EUV-capable observations would provide a critical foundation for interpreting potential biosignatures and assessing long-term habitability.  A step in this direction is the upcoming NASA MANTIS\footnote{\url{https://lasp.colorado.edu/missions/mantis/}} cubesat mission, scheduled to launch in 2027, which will provide low-resolution (30 \AA) EUV spectrophotometry of approximately 15–20 nearby active stars \citep{Indahl2022}. However, this sample will be limited in both size and diversity, and will not span the full range of ages and spectral types relevant to the broader HWO target list. 

Physical models of EUV emission based on stellar atmosphere codes remain limited (see \citealt{Linsky2024} for a more in-depth discussion). Codes capable of self-consistently modeling chromospheres, transition regions, and coronae exist for individual stars \citep[e.g., ][]{Peacock2019,Peacock2020, Tilipman2021}, but these models are not yet broadly applicable across stellar types and often do not capture the full extent of EUV-emitting layers. In lieu of these models, empirical scaling relationships are commonly used to estimate EUV fluxes from more readily observed diagnostics, such as X-rays and various FUV emission lines and continua, including \lya~\citep[e.g., ][]{Sanzforcada2011,Sanz-Forcada2025, Linsky2014,Chadney2015,king2018,France2018,Ketzer2023}. While these methods come with significant systematic uncertainties, particularly at low activity levels or for stars unlike the calibration sample \citep{france2022,zhang2022}, the relationships provide a pathway to estimate the unobservable EUV component for all high-priority HWO targets, especially when cross-calibrated and used in concert.

Understanding the XUV history of solar-type and low-mass stars is foundational for interpreting planetary atmospheric evolution. The combined effects of time-variable X-ray and EUV emission, especially during the first Gyr of stellar evolution, are believed to set the ``cosmic shoreline" for atmospheric retention \citep{Zahnle2017}. This framework is now being empirically tested for M dwarf planets using JWST \citep[e.g., ][]{Kreidberg2019,crossfield2022,Greene2023, Zieba2023, lustig2023, Moran2023, May2023, Kirk2024, Zhang2024}, but remains under-constrained for solar analogs. The Sun itself appears unusually quiet for its age and type \citep{Reinhold2020}, raising questions about how typical its radiation environment is for long-term atmospheric stability.

Filling in the missing EUV data (especially for young solar-type stars and older low-mass stars) will help establish a continuous picture of XUV evolution across stellar mass and age. Investment in completing the XUV inventory of HWO target stars, including reanalysis of archival data and acquisition of new observations where feasible, will provide a vital bridge between NASA's current exoplanet characterization efforts and the future science goals of HWO. Given the diagnostic richness of the FUSE bandpass (912--1187\,\AA) and its unique contribution to characterizing stellar UV environments, expanded observations in this wavelength range would also be highly valuable in complementing limited EUV datasets and refining stellar inputs to planetary atmosphere models.

\subsection{Future Observing Campaigns}\label{subsec:future}

Many of the existing high-energy observations of potential HWO targets were often motivated by prior knowledge of elevated activity levels or confirmed exoplanet hosts. As a result, a significant number of stars lack full spectral coverage across the high-energy domain, and only a small subset has been observed contemporaneously across multiple wavelength regimes (important for active stars, where high energy flux can vary by orders of magnitude over stellar cycles). This incomplete coverage presents a challenge for forward modeling of exoplanet atmospheres and stellar irradiation histories.

In the UV, HST remains the only practical facility capable of providing the high-resolution, high-sensitivity spectra required to characterize the majority of faint HWO-relevant targets  (Table \ref{tab:active}). In the X-ray, the recent suspension of eROSITA operations further underscores the need to rely on the capabilities of existing observatories (e.g., Chandra and XMM-Newton) for new observations. Given the uncertain operational timelines of both HST and Chandra, and the fact that no comparable UV or X-ray observatory is expected for at least 5–10 years (with UVEX offering more limited spectral resolution and wavelength coverage), there is a timely opportunity to design observing campaigns that can fill critical gaps in the high-energy datasets for these likely HWO target stars. Priority efforts include acquiring X-ray and UV spectra for stars with incomplete or outdated measurements, as well as obtaining contemporaneous multiwavelength data to study variability due to flares, rotation, and stellar cycles. These data are essential not only for constructing robust stellar models but also for propagating systematic uncertainties into future HWO-driven studies of planetary atmospheres.

The following observing efforts are feasible with currently available resources:
\begin{itemize}
\item X-ray flux measurements for the 24 target stars currently lacking reliable X-ray detections
\item X-ray spectra and updated fluxes for the 26 stars with only legacy ROSAT measurements (of any data quality)
\item FUV and/or NUV spectra for the 32 stars without existing reliable IUE or HST data in at least one band
\item FUV and/or NUV spectra for the 32 stars with poor quality IUE observations but no HST follow-up
\item HST FUV spectra for the 18 stars that currently have only HST NUV data (even if IUE FUV data exists), and HST NUV spectra for the 4 stars with only HST FUV data, in order to obtain a complete set of HST observations in both bands.
\item High resolution HST FUV and NUV LISM observations along specific sight lines (to enable a robust EUV and \lya reconstructions)
\item X-ray spectra for the 17 stars with usable FUV and NUV spectra, but lack X-ray spectra
\item Contemporaneous X-ray, FUV, and NUV fluxes for the majority of the targets
\end{itemize}

\begin{deluxetable}{lcc}[htb!]
\tablecaption{Active high energy missions.}
\label{tab:active}
\setlength{\tabcolsep}{4pt}
\tablehead{
\colhead{Mission Name} & \colhead{Energy/Wavelength Range} & \colhead{Potential Programs}
}
\startdata
XMM-Newton       & 0.2--8 keV                        & \makecell[c]{X-ray spectra,\\ spectral variability} \\
Chandra (ACIS-S) & 0.1--10 keV                       & \makecell[c]{X-ray spectra} \\
Chandra (HRC-I)  & 0.1--10 keV                       & \makecell[c]{X-ray fluxes,\\ flux variability} \\
eRosita          & 0.2--10 keV                       & \makecell[c]{X-ray spectra,\\ spectral variability} \\
NICER            & 0.2--12 keV                       & \makecell[c]{X-ray spectra} \\
Swift            & 0.2--10 keV / 170--650 nm        & \makecell[c]{X-ray and NUV fluxes,\\ flux variability} \\
Hubble (STIS)    & 1140--10000 \AA                      & \makecell[c]{FUV and NUV spectra,\\ spectral variability} \\
Hubble (COS)     & 1100--1900 \AA                       & \makecell[c]{FUV spectra,\\ spectral variability} \\
CUTE             & 2490--3310 \AA                       & \makecell[c]{NUV spectra} \\
\enddata
\end{deluxetable}

\begin{deluxetable}{lccccc}[htb!]
\tablecaption{Funded future high energy missions.}
\label{tab:future}
\setlength{\tabcolsep}{5pt}
\tablehead{
\colhead{\makecell[c]{Mission\\Name}} & \colhead{\makecell[c]{Wavelength\\Coverage}} & \colhead{\makecell[c]{Resolving\\Power}} & \colhead{\makecell[c]{Stage of\\Preparation}} & \colhead{\makecell[c]{GO\\Program?}} & \colhead{\makecell[c]{Potential\\Programs} }
}
\startdata
SPARCS	&	150 -- 250 nm	&	Photometry	&	\makecell[c]{Sched. launch\\Nov 2025?}	&	No  & \makecell[c]{FUV and NUV fluxes\\ and flux variability}	\\
MAUVE	&	200 -- 700 nm	&	R $\sim$20–65&	\makecell[c]{Sched. launch\\Oct 2025}	&	No	&	\makecell[c]{NUV fluxes\\ and flux variability}	\\
MANTIS	&	10 -- 600 nm	&	R $\sim$20–300	&	Est. launch 2027 &	No	&	\makecell[c]{Simultaneous EUV, FUV, NUV fluxes and spectra,\\ and flux variability} \\
UVEX	&	\makecell[c]{139 – 190 nm\\203 – 270 nm\\115 - 265 nm}	&	\makecell[c]{Photometry\\Photometry\\R $>$ 1000}	&	Est. Launch 2030	&	Yes	&	\makecell[c]{FUV and NUV fluxes\\ (potential FUV/NUV flux variability\\ and FUV spectra in GO)}	\\
AXIS	&	0.2 -- 10 keV	&	--	&	Est. Launch 2032	&	Yes	&	\makecell[c]{Low-resolution X-ray spectra,\\ spectral variability} \\
NewAthena & 0.2 -- 12 keV & R $\sim$100-1000 & Est. Launch 2037 & Yes & High- and low-resolution X-ray spectra \\
\enddata
\end{deluxetable}

Several CubeSat, SmallSat, and medium-class UV missions are on the horizon (Table \ref{tab:future}), though their long-term prospects remain uncertain pending the outcome of the federal appropriations process. While their effective areas and spectral capabilities are generally insufficient for the faint stars prioritized in HWO planning, a number of missions currently in build or Phase A have the potential to extend UV and X-ray coverage of likely HWO targets. Through a combination of survey modes, cadence strategies, and Guest Observer (GO) programs, these missions may provide updated flux measurements and valuable high-energy stellar context.

The Star-Planet Activity Research CubeSat (SPARCS\footnote{\url{https://sparcs.asu.edu/}}; \citealt{Ardila2018}) is a 6-U design that will obtain simultaneous FUV and NUV photometry for 20 M and K dwarfs over 1--3 full rotation periods. Although SPARCS lacks a GO program, its minute-cadence light curves will provide a homogeneous census of flare color, energy, and frequency as a function of age for low-mass stars.

Similarly, MAUVE\footnote{\url{https://bssl.space/mauve/}} will deliver low-resolution (2000–7000 \AA) UV spectrophotometry of the brightest, most active planet hosts, essentially replacing the time-domain capability of IUE and GALEX. MANTIS (Monitoring Activity of Nearby sTars with UV Imaging and Spectroscopy; \citealt{Indahl2022}) is a 16-U CubeSat designed for simultaneous FUV, NUV, and EUV monitoring, providing the first stellar EUV measurements since EUVE. Like SPARCS, MAUVE and MANTIS are missions without formal GO time, yet even their ``general release'' light curves and spectra will fill key gaps in flare energy distributions and quiescent spectra for bright, nearby stars.

Two larger Explorer and Probe class missions include GO allocations.
The NASA Explorer-class mission UVEX\footnote{\url{https://www.uvex.caltech.edu/}} (UltraViolet EXplorer; \citealt{Kulkarni2021}) combines a wide-field two-band imager with a multi-slit spectrograph to conduct a synoptic all-sky UV survey 50--100 times deeper than GALEX. Its GO program will enable pointed FUV spectroscopy, as well as multi-epoch FUV and NUV imaging of individual stars, offering both flux baselines and variability diagnostics for many target stars. UVEX is expected to launch in 2030. However, the mission faces an uncertain future given the proposed funding changes in the Fiscal Year 2026 President's Budget Request for NASA.

The NASA probe-class mission AXIS\footnote{\url{https://axis.umd.edu/}} (Advanced X-ray Imaging Satellite; \citealt{Reynolds2023}, currently in Phase A development) will provide arcsecond imaging resolution and high sensitivity from 0.3–-10\,keV. With a wide 24 $\times$ 24\,arcmin$^2$ FOV and fast slew time, the survey capabilities of AXIS will provide ample opportunity to capture the X-ray flux and variability of HWO targets of interest. AXIS will reserve 70\% of observing time for guest observer science, permitting both low-resolution spectroscopy and monitoring of target stars. The anticipated launch date is 2032.

Finally, the European Space Agency (ESA) flagship mission NewAthena\footnote{\url{https://www.esa.int/Science_Exploration/Space_Science/NewAthena_factsheet}} \citep{Cruise2025} will provide high resolution X-ray spectroscopy ($R \sim$ 100--1000) with the X-IFU instrument and wide-field low resolution spectroscopy with the Wide Field Imager (WFI). The imaging resolution of NewAthena will be comparable to that of XMM-Newton ($\sim$ 10$\arcsec$ point spread function), making it difficult to resolve binaries, but otherwise providing an opportunity to get high-signal and high-quality spectra from key target systems. The anticipated launch date is 2037.

\section{Conclusions}

In this paper, we compiled a multiwavelength dataset for the 98 highest priority stars in the TSS25 list, drawing on archival X-ray, EUV, FUV, and NUV observations. We also estimated NUV fluxes for all 13,000 stars in the TSS25 list, and estimated H~{\sc i} column densities toward all Tier 1 stars using a recent 3D map of the local interstellar medium. Together, these data enable estimates of the high-energy radiation environments for HWO target stars, providing essential inputs for modeling photochemistry, atmospheric loss, and surface UV conditions on potentially habitable exoplanets. This dataset forms a critical foundation for future efforts to evaluate planetary habitability and prioritize targets for spectroscopic follow-up with HWO.

The archival data analysis revealed substantial limitations in the current high-energy observational coverage of stars identified as high-priority HWO targets. The sample of stars with archival X-ray and UV observations is highly heterogeneous, with significant gaps in spectral completeness and data quality. In the X-ray regime, only 26\% of the sample has spectroscopic data, while 42\% are limited to photometric measurements. UV coverage is somewhat more extensive, with 64\% of targets having some spectroscopic observations; however, the majority of these come from the IUE archive. While IUE provides broad spectral access, it lacks sensitivity to key diagnostic lines, most notably \lya. Restricting the sample to stars observed with HST, only 31\% have spectroscopy in both the NUV and FUV.

Comprehensive high-energy spectral energy distributions (SEDs) are rare. Only 2\% of the stars have usable spectroscopic measurements in all four wavelength regimes (X-ray, EUV, FUV, NUV), and just 17\% have spectroscopy in X-ray, FUV, and NUV. Only 11\% of stars meet the full multiwavelength criteria required to construct detailed SEDs akin to those produced by programs such as MUSCLES \citep{France2016}, including high-resolution measurements of \lya. Moreover, most available datasets represent single-epoch observations, despite the fact that high-energy emission from low-mass stars is known to be strongly variable \citep[e.g.,][]{Loyd18a}. A detailed inventory of repeat observations was beyond the scope of this effort, but it is evident that very few stars have sufficient time-domain coverage to characterize variability or flare activity. The practice of combining non-simultaneous spectroscopy across multiple bands introduces additional uncertainties, and few targets currently have coordinated observations that would allow construction of a temporally consistent SED.

These findings underscore a critical need for dedicated precursor campaigns to obtain comprehensive, high-quality high-energy spectra for HWO target stars. Accurate models of photochemistry, atmospheric loss, and surface UV environments on potentially habitable exoplanets depend sensitively on the spectral shape and intensity of stellar X-ray through NUV radiation. However, such models are presently limited by the heterogeneous, incomplete, and often low-resolution nature of available data. In particular, just 40\% of Tier 1 targets have FUV spectroscopy, and only 35\% have FUV photometry of any kind. Only a small fraction of stars have measurements of \lya or reconstructed EUV fluxes, despite their critical role in driving upper atmospheric escape. In many cases, the best available X-ray data come from shallow ROSAT detections or are entirely missing.

To address these limitations, targeted efforts (like those listed in Section \ref{subsec:future}) using existing facilities, particularly Chandra, XMM-Newton, and HST, can significantly improve the high-energy dataset. In parallel, the development of validated empirical or semi-empirical models, including improved EUV reconstructions, forward models, and flare-informed variability prescriptions, is essential. Ultimately, achieving the data quality and coverage required for robust atmospheric characterization will likely require a new UV or EUV mission. These preparatory steps are vital not only for interpreting future exoplanet spectra but also for identifying and prioritizing the most promising habitable-zone targets for observation with HWO.

\begin{acknowledgments}
This publication is a direct product of the HWO Target Stars and Systems sub-Working Group. The results and conclusions benefited greatly from group-wide discussions and analysis, as well as specific input from the Catalogs \& Databases Task Group. S.P. acknowledges support from NASA under award number 80GSFC24M0006. 
Part of this research was carried out at the Jet Propulsion Laboratory, California Institute of Technology, under a contract with the National Aeronautics and Space Administration (80NM0018D0004). 
\end{acknowledgments}


\facilities{IUE (SWP, LWP, LWR), ROSAT, HST (COS, STIS, GHRS), EUVE, FUSE, Chandra, GALEX, Swift (XRT, UVOT), eROSITA}

\software{\texttt{LISM\_NHI} \citep{Youngblood2025}}

\appendix

\begin{figure}[ht!]
    \centering
    \includegraphics[width=0.8\linewidth]{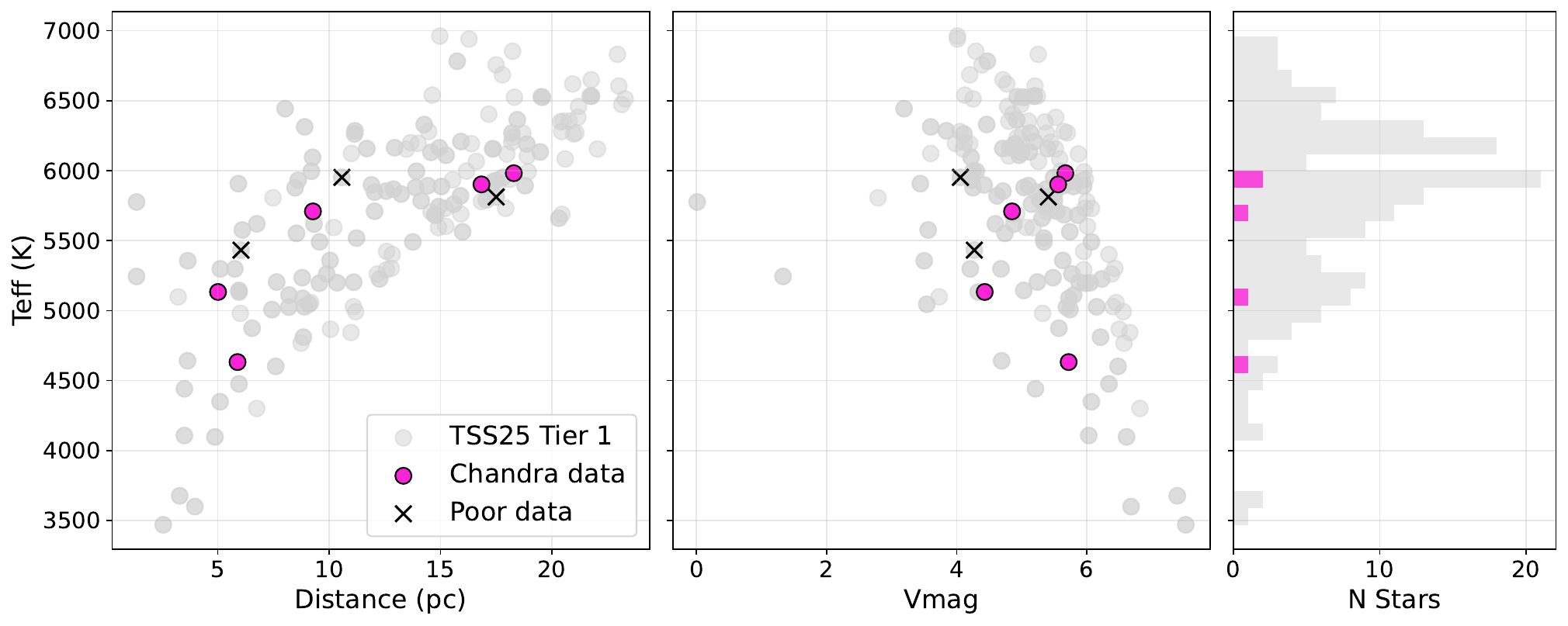}
    \includegraphics[width=0.8\linewidth]{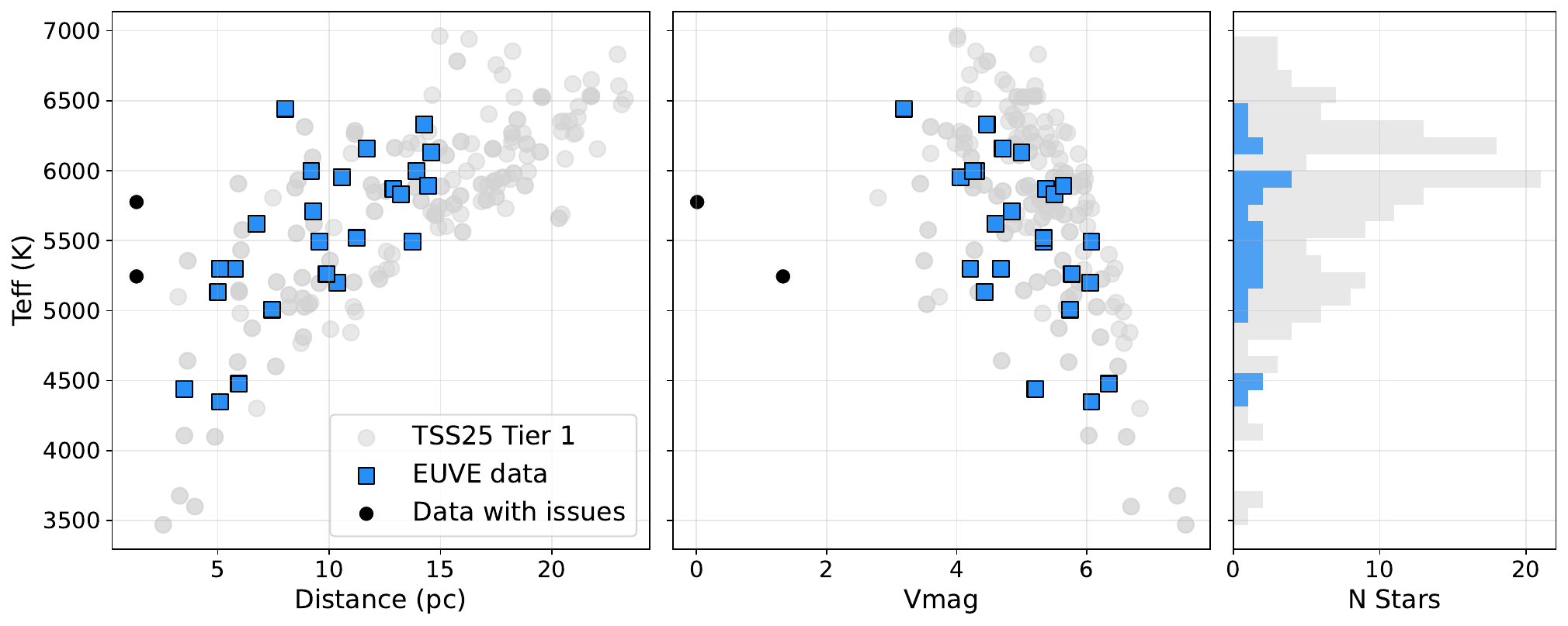}
    \includegraphics[width=0.8\linewidth]{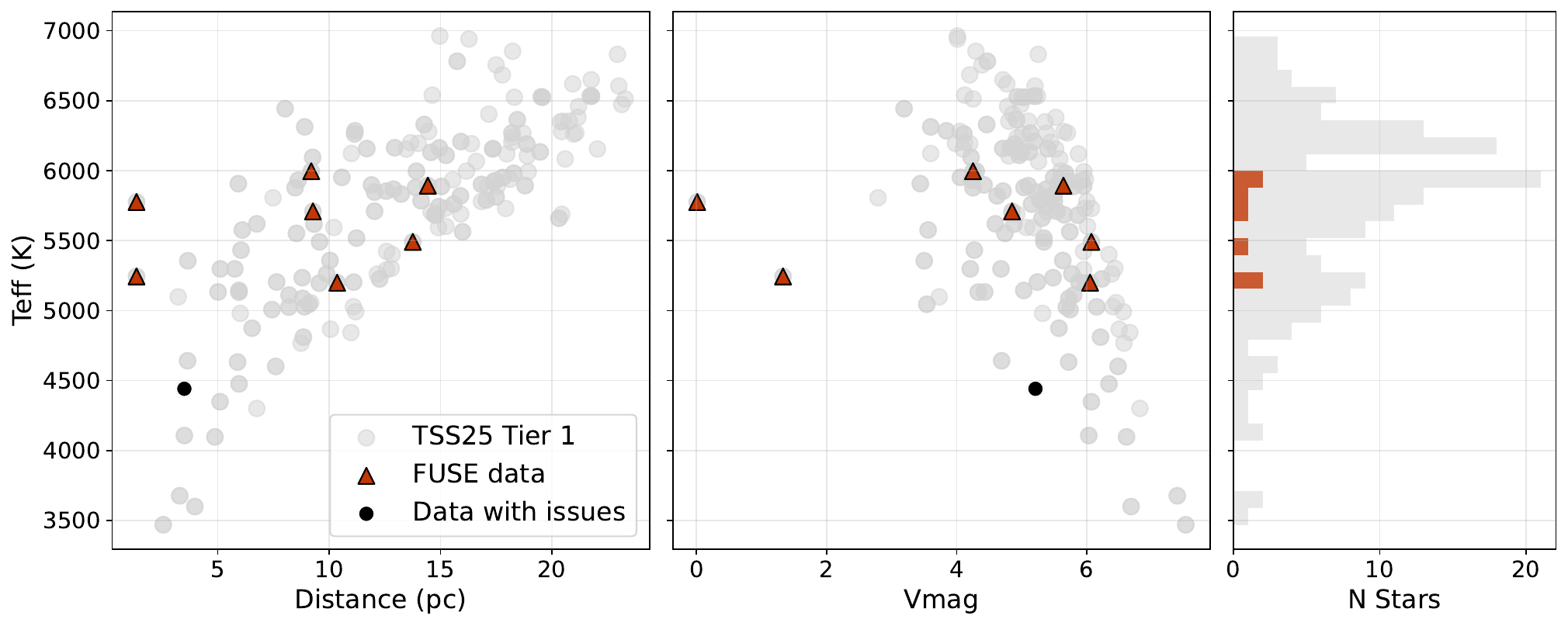}
    \caption{Same as Figure \ref{fig:full_demo}, but showing each observatory individually. \textit{Top:} Demographics of stars with observations from Chandra. \textit{Middle:} EUVE. \textit{Bottom:} FUSE. In each panel, $T_{\mathrm{eff}}$ is plotted against distance (left), $V$-band magnitude (middle), and the number of stars per $T_{\mathrm{eff}}$ bin (right). Histogram bars begin at zero and only count usable-quality data. usable-quality data are shown in color with the corresponding observatory symbol. Observations with known issues (e.g., unresolved binaries or upper limits) are shown as black filled circles. Poor-quality data or failed observations are indicated with black Xs.}
    \label{fig:a1}
\end{figure}

\begin{figure}[!htb]
    \centering
    \includegraphics[width=0.8\linewidth]{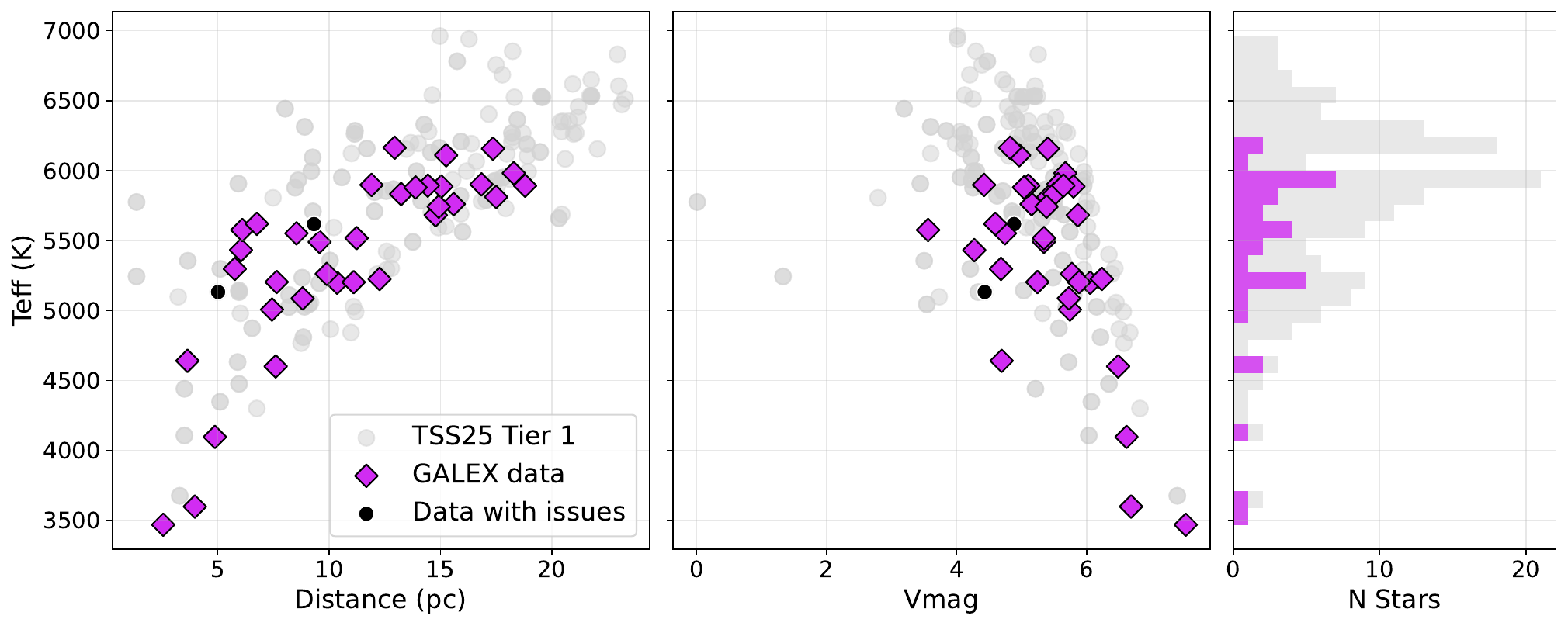}
    \includegraphics[width=0.8\linewidth]{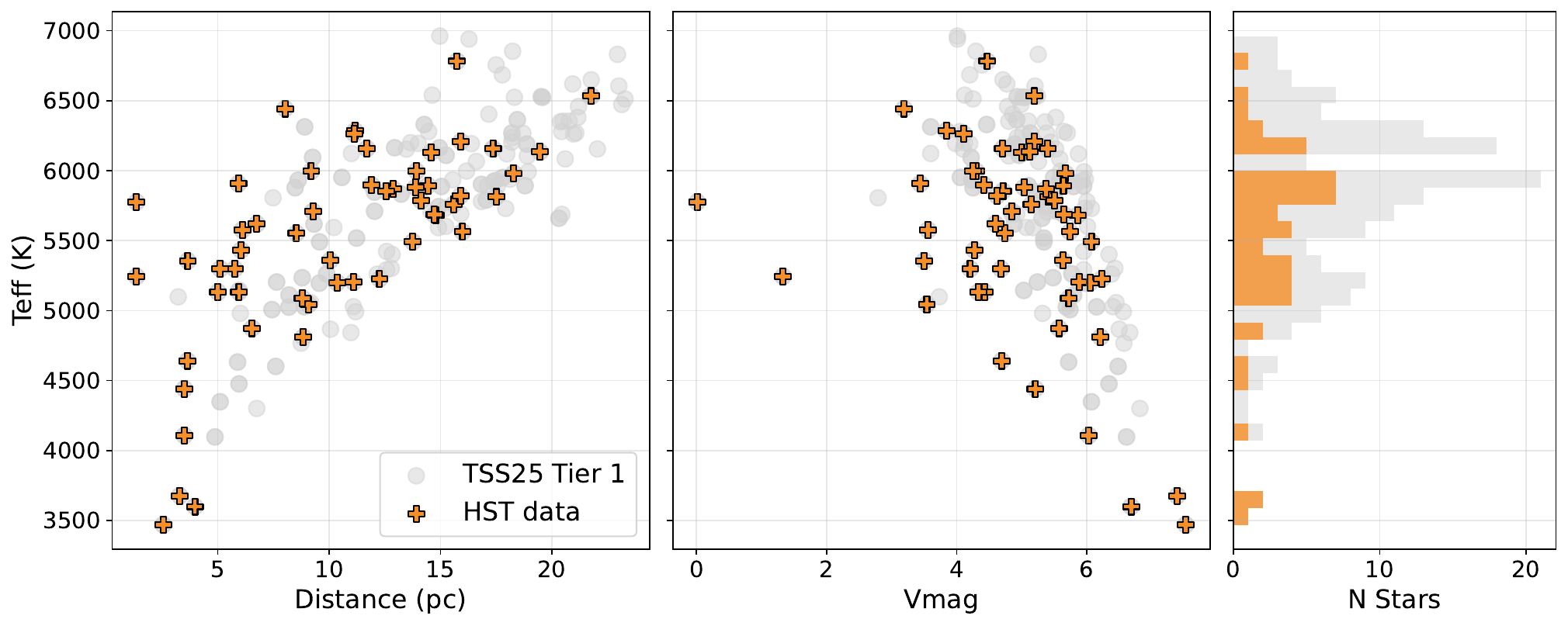}
    \includegraphics[width=0.8\linewidth]{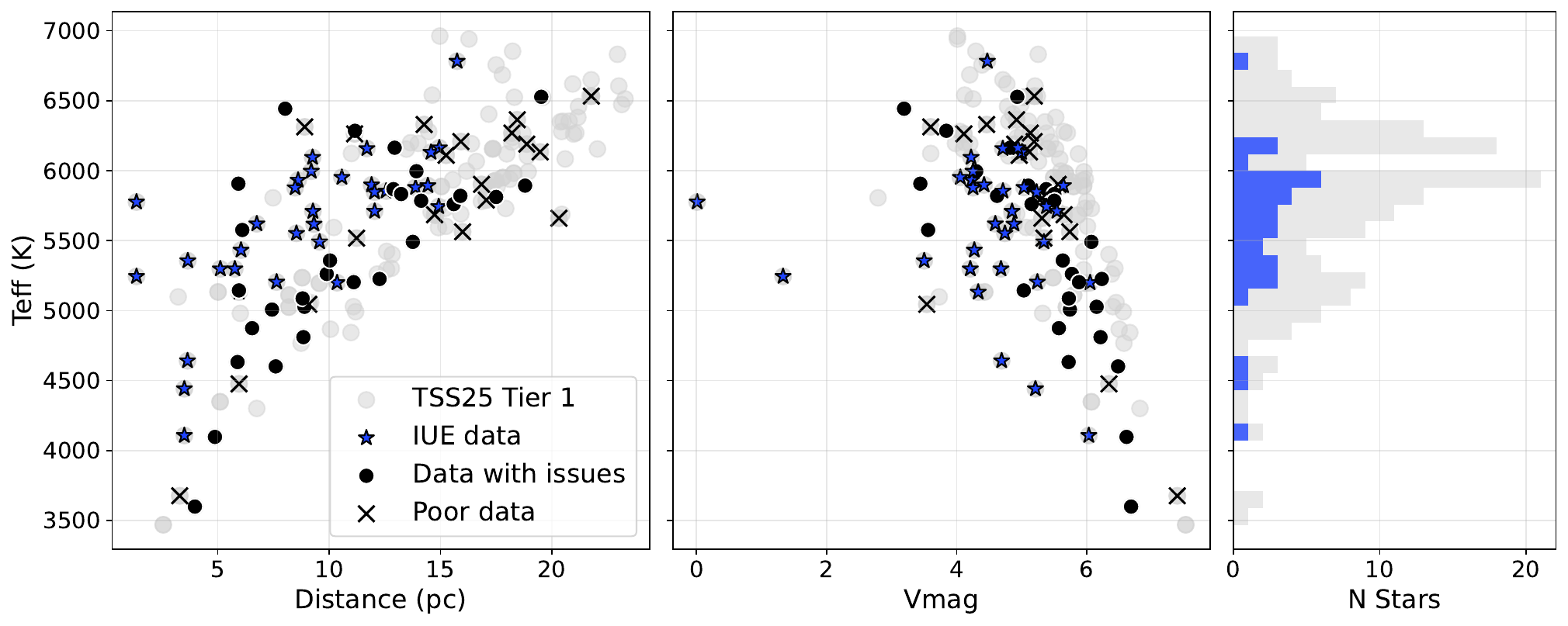}
    \caption{Same as Figure \ref{fig:a1}. \textit{Top:} GALEX, \textit{Middle:} HST, \textit{Bottom:} IUE.}
    \label{fig:a2}
\end{figure}

\begin{figure}[!htb]
    \centering
    \includegraphics[width=0.8\linewidth]{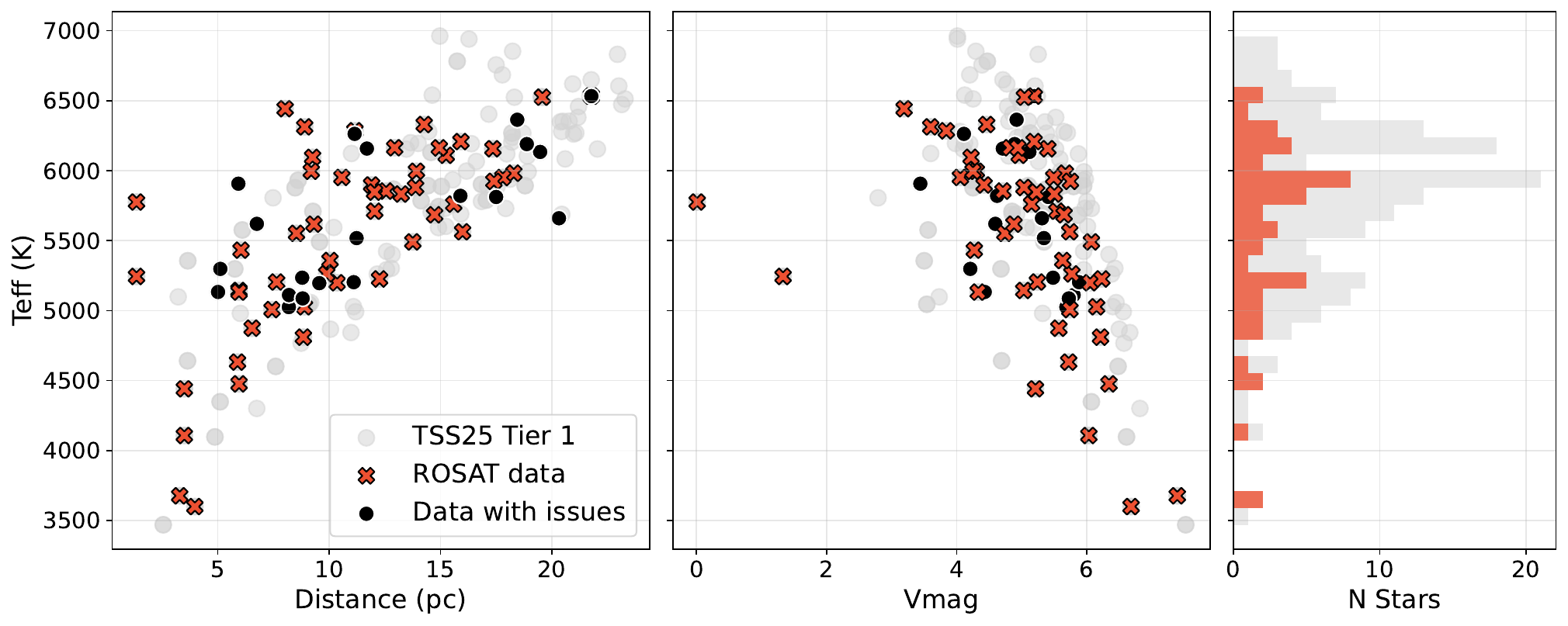}
    \includegraphics[width=0.8\linewidth]{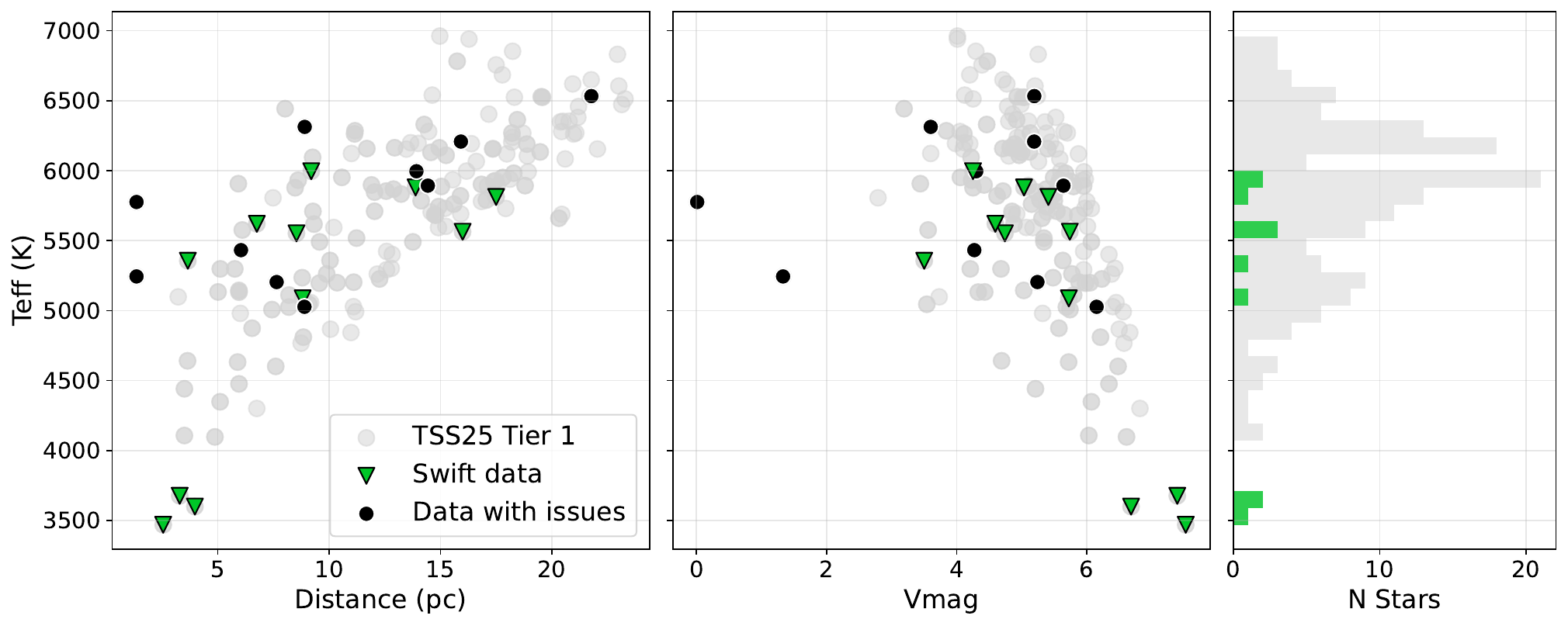}
    \includegraphics[width=0.8\linewidth]{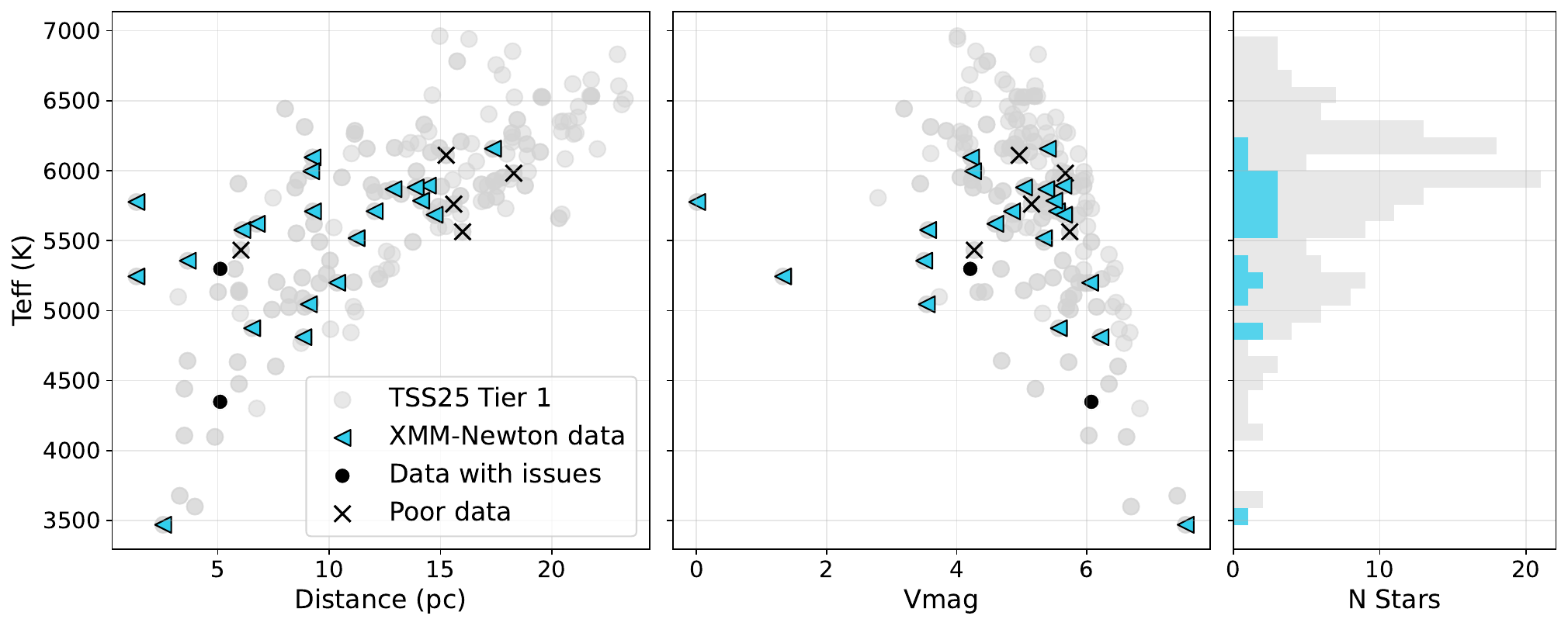}
    \caption{Same as Figure \ref{fig:a1}. \textit{Top:} ROSAT, \textit{Middle:} Swift, \textit{Bottom:} XMM-Newton.}
    \label{fig:a3}
\end{figure}

\begin{figure}[!htb]
    \centering
    \includegraphics[width=0.8\linewidth]{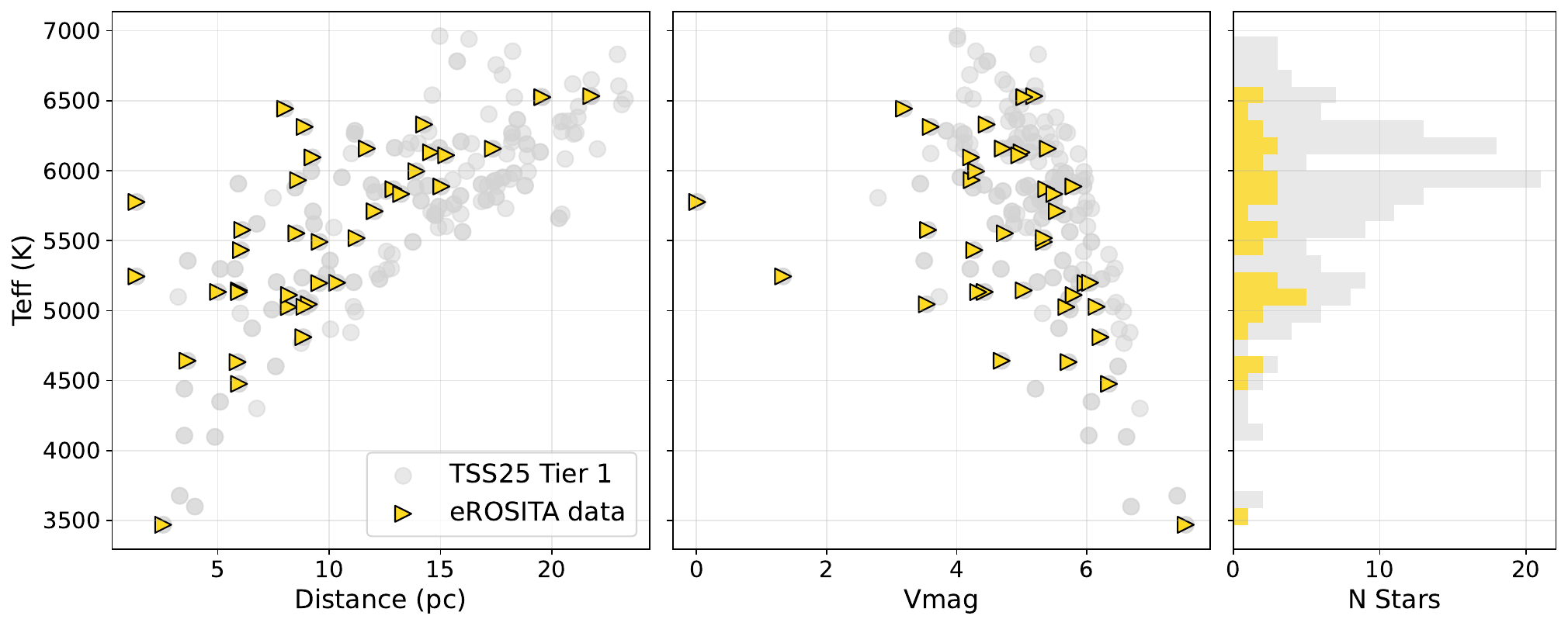}
    \caption{Same as Figure \ref{fig:a1}, for eROSITA.}
    \label{fig:a4}
\end{figure}

Figures \ref{fig:a1}, \ref{fig:a2}, \ref{fig:a3}, and \ref{fig:a4} present demographic comparisons between the full TSS25 Tier 1 target list and the subset of stars observed by each individual observatory. These figures also include markings for stars with flagged and failed observations.

\end{document}